\documentclass[letterpaper]{article} 
\usepackage{aaai23-r2hcai}  
\usepackage{times}  
\usepackage{helvet}  
\usepackage{courier}  
\usepackage[hyphens]{url}  
\usepackage{graphicx} 
\usepackage{natbib}  
\usepackage{caption} 
\usepackage{algorithm}
\usepackage{algorithmic}
\usepackage{newfloat}
\usepackage{listings}
\DeclareCaptionStyle{ruled}{labelfont=normalfont,labelsep=colon,strut=off} 
\floatstyle{ruled}
\newfloat{listing}{tb}{lst}{}
\floatname{listing}{Listing}
\usepackage{blkarray,arydshln,lscape,comment,bbold,authblk, amsmath,mathtools,cuted}
\usepackage{etoolbox}
\AfterEndEnvironment{strip}{\leavevmode}
\makeatletter
  \renewcommand*\env@matrix[1][*\c@MaxMatrixCols c]{%
    \hskip -\arraycolsep
    \let\@ifnextchar\new@ifnextchar
  \array{#1}}
\makeatother
\newcommand\scalemath[2]{\scalebox{#1}{\mbox{\ensuremath{\displaystyle #2}}}}
\newcommand\numberthis{\addtocounter{equation}{1}\tag{\theequation}}

\usepackage{fancyhdr}
\fancypagestyle{firstpagehf}
{
   \fancyhf{}
   \fancyhead[HC]{Evaluation of wait-time-saving effectiveness of triage algorithms}
   \fancyfoot[C]{\thepage}
}

\title{Evaluation of wait-time-saving effectiveness of triage algorithms}
\author{
    Yee Lam Elim Thompson$^1$ \hspace{0.25em}
    Gary M Levine$^1$ \hspace{0.25em}
    Weijie Chen$^1$ \hspace{0.25em}
    Berkman Sahiner$^1$ \hspace{0.25em}
    Qin Li$^1$ \hspace{0.25em} 
    Nicholas Petrick$^1$ \hspace{0.25em}
    Jana G Delfino$^1$ \hspace{0.25em}
    Miguel A Lago$^1$ \hspace{0.25em}
    Qian Cao$^1$ \hspace{0.25em}
    Qin Li$^1$\footnote{Dr. Qin Li has left the FDA and is currently the director of Translational Medicine at Astrazeneca. Her contribution to this work was made when she was at the FDA.} \hspace{0.25em}
    Frank W Samuelson$^1$
}
\affiliations {
    \textsuperscript{\rm 1} The U.S. Food and Drug Administration \quad
    YeeLamElim.Thompson@fda.hhs.gov
}

\usepackage{bibentry}

\begin{document}
\thispagestyle{firstpagehf}
\maketitle

\begin{abstract}
In the past decade, Artificial Intelligence (AI) algorithms have made promising impacts to transform healthcare in all aspects. One application is to triage patients’ radiological medical images based on the algorithm's binary outputs. Such AI-based prioritization software is known as computer-aided triage and notification (CADt). Their main benefit is to speed up radiological review of images with time-sensitive findings. However, as CADt devices become more common in clinical workflows, there is still a lack of quantitative methods to evaluate a device’s effectiveness in saving patients' waiting times. In this paper, we present a mathematical framework based on queueing theory to calculate the average waiting time per patient image before and after a CADt device is used. We study four workflow models with multiple radiologists (servers) and priority classes for a range of AI diagnostic performance, radiologist's reading rates, and patient image (customer) arrival rates. Due to model complexity, an approximation method known as the Recursive Dimensionality Reduction technique is applied. We define a performance metric to measure the device’s time-saving effectiveness. A software tool is developed to simulate clinical workflow of image review/interpretation, to verify theoretical results, and to provide confidence intervals  of the performance metric we defined. It is shown quantitatively that a triage device is more effective in a busy, short-staffed setting, which is consistent with our clinical intuition and simulation results. Although this work is motivated by the need for evaluating CADt devices, the framework we present in this paper can be applied to any algorithm that prioritizes customers based on its binary outputs.    
\end{abstract}

\section{Introduction}
\label{sec:intro}

The fast-growing development of artificial intelligence (AI) and machine learning (ML) technologies bring a potential to transform healthcare in many ways.
One emerging area is the use of AI/ML as Software as a Medical Device (SaMD) in radiological imaging to triage patient images with time-sensitive findings for image interpretation \citep{VanLeeuwen2021}.
These devices are known as computer-aided triage and notification (CADt) devices, by which medical images labeled as positive by an AI algorithm are prioritized in the radiologist's reading queue.
The major benefit of a CADt device is to increase the likelihood of timely diagnosis and treatment of severe and time-critical diseases such as large vessel occlusion (LVO), intracranial hemorrhage (ICH), pneumothorax, etc.
In 2018, the U.S. Food and Drug Administration (FDA) granted marketing authorization to the first CADt device for potential LVO stroke patients via the \textit{de novo} pathway \citep{FDANewsRelease20180213}. 
Since then, multiple studies have shown improvements in patient treatment and clinical outcomes due to the use of CADt devices \citep{Hassan2020, YahavDorat2021, Rodrigues2019, Hassan2021}.
Most of these analyses focus on the diagnostic performance when evaluating these CADt devices, but a quantitative estimate of time savings for truly diseased (signal-present) patient images in a clinical environment remain unclear.
Therefore, the goal of this work is to fill this gap by developing a queueing-theory based tool to characterize the time-saving effectiveness of a CADt device in a given clinical setting.

\begin{figure*}[t]
\centering
\includegraphics[width=0.85\textwidth]{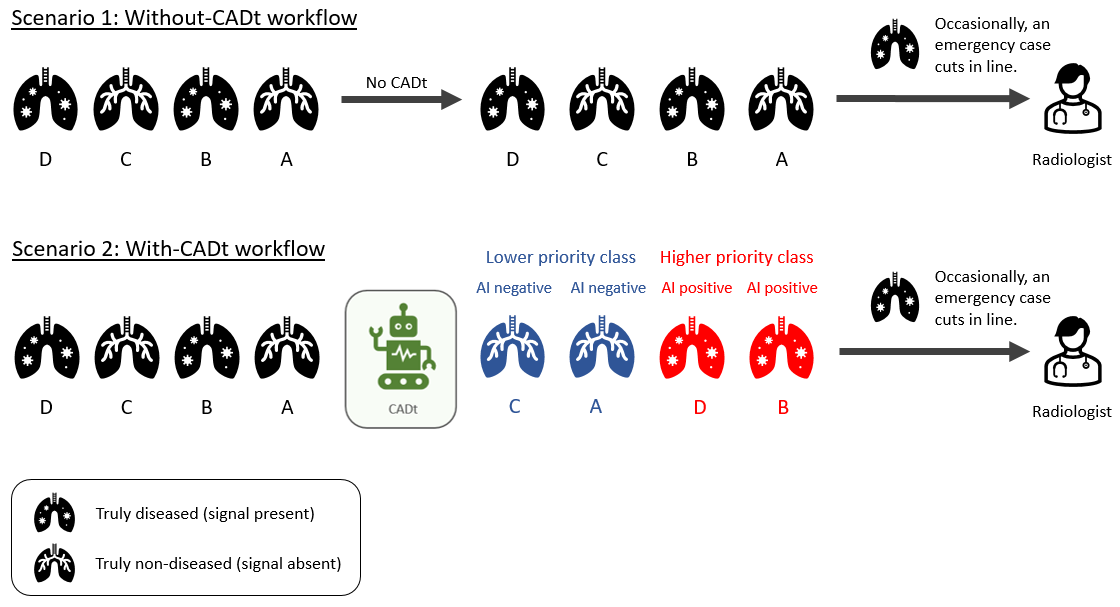}
\caption{Radiologist workflows without and with a CADt device.
\textit{Top}: the without-CADt scenario in which patient images are reviewed in the order of their arrival.
\textit{Bottom}: the with-CADt workflow in which AI-positive patient images are reviewed first before the AI-negative images.
In both scenarios, the radiologist may be interrupted by emergent cases.
All cartoon icons are adopted from Microsoft PowerPoint application.
}
\label{fig:intro_WithWithoutCADtWorkflow}
\end{figure*}

Figure \ref{fig:intro_WithWithoutCADtWorkflow} illustrates the radiologist workflows without and with a CADt device being used.
In the standard of care without a CADt device, patient images are reviewed by a radiologist on a first-in, first-out (FIFO) basis.
In the context of queuing theory, our servers are radiologists, and our customers are patient images.
Occasionally, the radiologist may be interrupted by an emergent case, for example, when a physician requests an immediate review of a specific patient image.
To distinguish these emergent cases from those in the reading queue, we call the images in the reading list ``non-emergent."
If a CADt device is included in the workflow, the device only analyzes non-emergent patient images.
Cases labeled as AI-positive are either flagged or moved up in a radiologist's reading list, giving them higher priority, and the radiologist will review those cases before all AI-negative patient images.
Just like the without-CADt scenario, the radiologist may be interrupted by emergent cases, which always have the highest priority over other images.
Overall, without a CADt, we have a queue with two priority classes, and we have a queue with three priority classes in a with-CADt scenario.

It is noted that, though applied to radiology clinics, the mathematical frameworks presented here could be used to evaluate discrimination algorithms in other queueing contexts.
For example, algorithms may attempt to identify customers or jobs who may require less service time and place them into a higher priority class, thereby reducing overall wait time for customers on average.

\section{Parameters}
\label{sec:parameters}

Before applying queueing theory, a few parameters are defined to describe the clinical setting.
\begin{itemize}
  \item $f_{\text{em}}$ is the fraction of emergent patient images with respect to all patient images.
  \item $\lambda$ is the Poisson arrival rate of all patient images.
  Patient images can be divided into subgroups, and each subgroup $i$ has a Poisson arrival rate $\lambda_i = p_i\lambda$, where $p_i$ is the fraction of image subgroup $i$ with respect to all patient images.
  \item The disease prevalence $\pi$ is defined within the non-emergent patient population, i.e.
  \begin{equation}\label{eqn:pi_textdef}
      \pi = \frac{\text{Number of diseased, non-emergent cases}}{\text{Number of non-emergent cases}}.
  \end{equation}
  \item CADt diagnostic performance is defined by its sensitivity ($\text{Se}$) and specificity ($\text{Sp}$), which are also defined within the non-emergent patient images i.e.
  \begin{equation*}
      \text{Se} = \scalemath{0.85}{\frac{\text{Number of AI-positive, diseased, non-emergent cases}}{\text{Number of diseased, non-emergent cases}}},
  \end{equation*}
  and
  \begin{equation*} \scalemath{0.85}{
      \text{Sp} = \frac{\text{Number of AI-negative, non-diseased, non-emergent cases}}{\text{Number of non-diseased, non-emergent cases}}}.
  \end{equation*}
  \item $N_{\text{rad}}$ is the number of radiologists on-site.
  Typically, a clinic has at least one radiologist at all times. 
  For a larger hospital, multiple radiologists may be available during the day.
  \item The radiologist's reading rates are denoted by $\mu$'s.
  For emergent (highest-priority) cases, the reading time $T_{\text{em}}$ is assumed to be exponentially distributed with an average reading rate $\mu_{\text{em}}=1/\overline{T}_{\text{em}}$.
  For a non-emergent image, the average reading rate depends on the radiologist's diagnosis i.e. $\mu_D$ if diseased image or $\mu_{ND}$ if non-diseased image.
  Therefore, in the without-CADt scenario, the reading time of the non-emergent (lower-priority) cases follows a hyperexponential distribution where the mean $(1/\mu_{\text{nonEm}})$ is determined by the mean reading rates of the two subgroups and the probability of disease prevalence $\pi$.
  \begin{equation}\label{eqn:muNonEm}
    \frac{1}{\mu_{\text{nonEm}}} = \frac{\pi}{\mu_D} + \frac{1-\pi}{\mu_{ND}}.
  \end{equation}
  In the with-CADt scenario, the average reading rates for AI-positive (middle-priority) and AI-negative (lowest-priority) classes are denoted by $\mu_+$ and $\mu_-$ respectively.
  The AI-positive group consists of true-positive (TP) and false-positive (FP) patients, and the probability that an AI-positive case is a TP is defined by the positive predictive value (PPV).
  Hence, 
  \begin{equation}\label{eqn:muPlus_eff}
    \frac{1}{\mu_+} = \frac{\text{PPV}}{\mu_D} + \frac{1-\text{PPV}}{\mu_{ND}}.
  \end{equation}  
  Similarly, the average AI-negative reading rate is given by
  \begin{equation}\label{eqn:muMinus_eff}
    \frac{1}{\mu_-} = \frac{1-\text{NPV}}{\mu_D} + \frac{\text{NPV}}{\mu_{ND}},
  \end{equation}  
  where $\text{NPV}$ is the probability that an AI-negative case is a true-negative (TN).
  \item $\rho$ is the traffic intensity defined as $\rho = \lambda/\mu_{\text{eff}}$, where  $\mu_{\text{eff}}$ is effective reading rate considering all priority classes and $N_{\text{rad}}$ in the queueing system.
  $\rho$ ranges from 0 with no patient images arriving to 1 implying a very congested hospital.
  \item With regard to the queueing discipline, when no CADt device is used, patient images are read in the order of their arrival time i.e. first-in first-out (FIFO).
  In the with-CADt scenario, we consider a preemptive-resume priority scheduling:
  whether or not a CADt device is used, whenever a higher-priority patient image enters the system, the reading of a lower-priority patient image will be interrupted and later resumed.
  Although in reality some radiologists may prefer finishing up the current lower-priority image when a CADt device flags a higher-priority case (which would be a non-preemptive-resume priority), many CADt devices are designed assuming a radiologist reads the flagged cases immediately.
  Therefore, a preemptive-resume priority is assumed in this work.
\end{itemize}

To assess the time-saving effectiveness of a given CADt device in a clinical setting defined by the above parameters, we first define four radiologist workflow models in Section \ref{sec:models}. 
For each of the models, we provide the Markov chain matrices to compute the mean waiting time for each priority class in both with- and without-CADt scenarios.
Section \ref{sec:simulation} discusses an in-house simulation software developed to verify theoretical results and to provide confidence intervals around the theoretical mean time savings.
Section \ref{sec:metric} defines a metric that quantifies the time-saving effectiveness of a CADt device, and Section \ref{sec:results} discusses the results obtained from theory and simulation.

\section{Radiologist workflow models}
\label{sec:models}

We consider four radiologist workflow models:
\begin{itemize}
    \item Model A: The baseline model ($N_{\text{rad}} = 1$, $f_{\text{em}} = 0$, and $\mu_D = \mu_{ND}$)
    \item Model B: Model A but with emergent patient images ($N_{\text{rad}} = 1$, $f_{\text{em}} > 0$, and $\mu_D = \mu_{ND}$)
    \item Model C: Model B but with two radiologists  ($N_{\text{rad}} = 2$, $f_{\text{em}} > 0$, and $\mu_D = \mu_{ND}$)
    \item Model D: Model B but with different reading rates for diseased and non-diseased images ($N_{\text{rad}} = 1$, $f_{\text{em}} > 0$, and $\mu_D \neq \mu_{ND}$)
\end{itemize}

For each model, two calculations are performed: one assumes a without-CADt scenario, and the other assumes the use of a CADt device.
Each scenario has a set of \textit{states} that keeps track of the numbers of patient images in different priority classes.
The transition rates among states form a stochastic Markov chain matrix, from which the matrix geometric method is applied to calculate the set of state probabilities \citep{Stewart2009} .
For models involving multiple radiologists and priority classes, we apply the Recursive Dimensionality Reduction (RDR) method proposed by \cite{HarcholBalter2005} to facilitate the calculation.
Little's Law \citep{Stewart2009} is then applied to calculate the mean waiting time per patient image for each priority class involved.

\subsection{Model A: Baseline model}
\label{sec:queue_modelA}
We start with a simple model with the absence of emergent patient images ($f_{\text{em}} = 0$), one radiologist on-site ($N_{\text{rad}} = 1$), and identical reading rates for diseased and non-diseased subgroups ($\mu_D = \mu_{ND}$).

\subsubsection{Model A in without-CADt scenario}
\label{sec:queue_modelA_noCADt}

\begin{figure*}[t]
\centering
\includegraphics[width=0.9\textwidth]{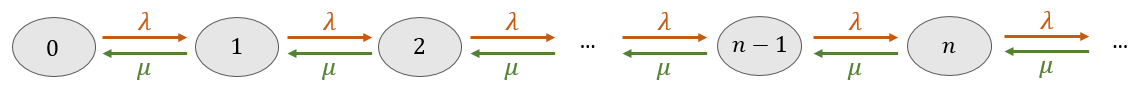}
\caption{The Markov chain transition diagram for non-emergent patient images in Model A without a CADt device.
     Gray bubbles represent the state $n_{\text{nonEm}}$, the total number of non-emergent patient images in the system.
     Top orange arrows represent the transition rate $\lambda$ to increase $n_{\text{nonEm}}$ one at a time, and bottom green arrows represent the transition rate $\mu$ to decrease $n_{\text{nonEm}}$ one at a time.}
\label{fig:simpleModel_transitionDiagram_noCADt}
\end{figure*}

First, we consider the without-CADt scenario.
Given that $f_{\text{em}} = 0$, only one priority class (the non-emergent subgroup) exists, and the arrival rate $\lambda$ is the arrival rate of non-emergent patient images $\lambda_{\text{nonEm}}$.
When $\mu_D = \mu_{ND}$ and with only 1 radiologist on-site, the effective reading rate for the non-emergent subgroup is $\mu_{\text{nonEm}} = \mu_D = \mu_{ND}$.
Hence, Model A turns into a classic M/M/1/FIFO queueing model \citep{Stewart2009}.
Its transition diagram is shown in Figure \ref{fig:simpleModel_transitionDiagram_noCADt}, from which the state probability $p_n$ is given by
\begin{equation}\label{eqn:stateProb_noCADt}
    p_{n_{\text{nonEm}}} = \rho_{\text{nonEm}}^n (1-\rho_{\text{nonEm}}),
\end{equation}
where $n_{\text{nonEm}}$ denotes the number of non-emergent patient image in the system.
From the state probability $p_{n_{\text{nonEm}}}$, the average waiting time per non-emergent patient image can be calculated by the following steps.
\begin{enumerate}
  \item Calculate the average number of non-emergent patient images in the system, $L$, from the state probability $p_{n_{\text{nonEm}}}$. That is, $L =\langle p_{n_{\text{nonEm}}} \rangle$, where $\langle \rangle$ is the expectation operator.
  \item Calculate the average response time per non-emergent patient image, $W$, via Little's Law i.e. $W = L/\lambda_{\text{nonEm}}$.
  \item Calculate the average waiting time in the queue per non-emergent patient, $W_{q_{\text{nonEm}}}$.
  Because $W$ is the sum of $W_{q_{\text{nonEm}}}$ and the mean radiologist's reading time $\overline{T}=1/\mu_{\text{nonEm}}$, we have $W_{q_{\text{nonEm}}}  = W - 1/\mu_{\text{nonEm}}$.
\end{enumerate}
In summary, the average waiting time per non-emergent patient image $W_{q_{\text{nonEm}}}$ in a without-CADt scenario is given by
\begin{equation}\label{eqn:Wq_noCADt}
    W_{q_{\text{nonEm}}} = \langle p_{n_{\text{nonEm}}}\rangle/\lambda_{\text{nonEm}} - 1/\mu_{\text{nonEm}}.
\end{equation}

\subsubsection{Model A in with-CADt scenario}
\label{sec:queue_modelA_CADt}

\begin{figure*}[t]
\centering
\includegraphics[width=0.7\textwidth]{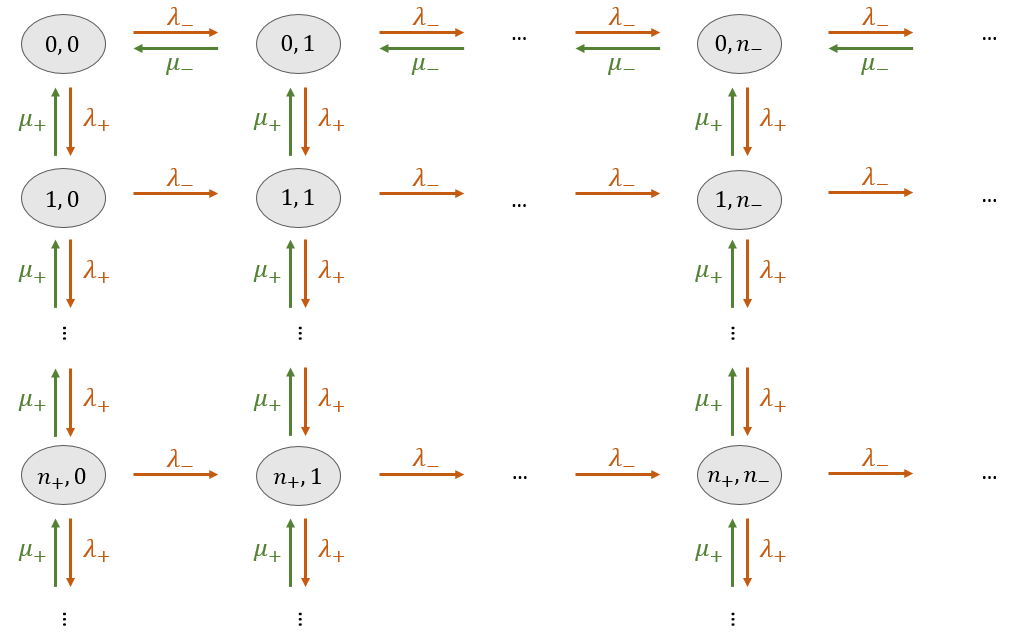}
\caption{The exact transition diagram for Model A in a with-CADt scenario.
     Gray bubbles represent the state $(n_+, n_-)$ defined by the numbers of AI-positive patient images, $n_+$, and AI-negative patient images, $n_-$, in the system.
     Each row represents the transition of increasing or decreasing $n_-$, and each column represents the transition of $n_+$.
     Note that AI-negative patient images can leave the system only when $n_+ = 0$, and hence the $\mu_-$ arrows only show up in the first row of transition diagram.}
\label{fig:simpleModel_transitionDiagram_low} 
\end{figure*}

When a CADt-device is used with no emergent patient images ($f_{\text{em}} = 0$), two priority classes exist: an AI-positive, higher-priority class and an AI-negative, lower-priority class.
The arrival rates of AI-positive and AI-negative classes depend on the CADt diagnostic performance.
\begin{equation}\label{eqn:lambdaPlus}
    \lambda_{+} = \big[\pi\text{Se} + (1-\pi)(1-\text{Sp})\big ]\lambda,
\end{equation}
\begin{equation}\label{eqn:lambdaMinus}
    \lambda_{-} = \big[\pi(1-\text{Se}) + (1-\pi)\text{Sp}\big]\lambda.
\end{equation}
The state of a two-priority class system is defined by the number of AI-positive cases $n_+$ and that of AI-negative $n_-$.
As shown in Figure \ref{fig:simpleModel_transitionDiagram_low}, the exact transition diagram is infinite in both horizontal ($n_-$) and vertical ($n_+$) directions.
With an assumed preemptive-resume priority scheduling, this 2D-infinity problem can be resolved using the Recursive Dimensionality Reduction (RDR) method \citep{HarcholBalter2005}, in which the tangled two-priority-class system is broken down into two independent calculations, one for each priority class.

First, we focus on the AI-positive, higher-priority system.
Because of the preemptive-resume queueing discipline, the AI-positive subgroup is not affected by the AI-negative images at all and is, by itself, a classic M/M/1/FIFO queueing model.
Therefore, to solve for the average waiting time per AI-positive patient image, one can reuse Figure \ref{fig:simpleModel_transitionDiagram_noCADt} and replace $n_{\text{nonEm}}$ by $n_+$.
The state probability for AI-positive patient images is modified based on Equation \ref{eqn:stateProb_noCADt}; 
\begin{equation}\label{eqn:stateProb_high_CADt}
    p_{n_+} = \rho_+^{n_+} (1-\rho_+),
\end{equation}
where $\rho_+ \equiv \lambda_+/\mu_+$ is the traffic intensity for the AI-positive subgroup only.
Following the steps in Equation \ref{eqn:Wq_noCADt}, the average waiting time per AI-positive patient image $W_{q_+}$ is given by
\begin{equation}\label{eqn:Wq_high_CADt}
    W_{q_+} = \langle p_{n_+}\rangle/\lambda_+ - 1/\mu_+.
\end{equation}

\begin{figure*}[t]
\centering
\includegraphics[width=0.75\textwidth]{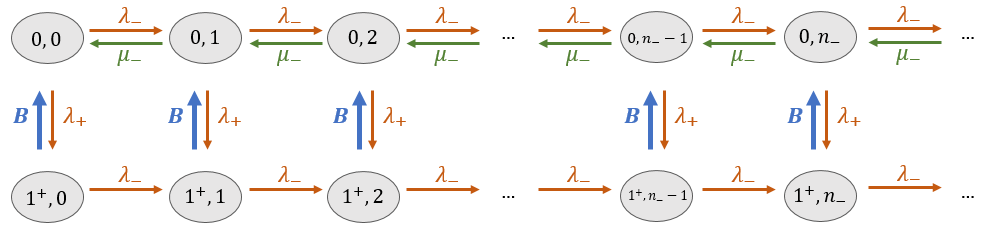}
\caption{The RDR-truncated transition diagram for AI-negative, low-priority patient images in Model A in the with-CADt scenario.
For every column in Figure \ref{fig:simpleModel_transitionDiagram_low}, all states beyond $n_+ \geq 1$ are truncated as $(1_+, n_-)$ with an approximated transition rate $\textbf{B}$.}
\label{fig:simpleModel_transitionDiagram_lowHBO} 
\end{figure*}

For the calculation of the AI-negative, lower-priority class, we cannot ignore the presence of AI-positive cases.
However, with only one radiologist, no AI-negative patient image can exit the system when $n_+ \geq 1$.
As noted by \cite{HarcholBalter2005}, there is no need to keep track of every state beyond $n_+ \geq 1$.
Hence, every column in Figure \ref{fig:simpleModel_transitionDiagram_low} can be truncated such that all states beyond $n_+ \geq 1$ are represented by $(1^+, n_-)$.
The RDR-truncated transition diagram is shown in Figure \ref{fig:simpleModel_transitionDiagram_lowHBO}.

\begin{figure}[t]
\centering
\includegraphics[width=\columnwidth]{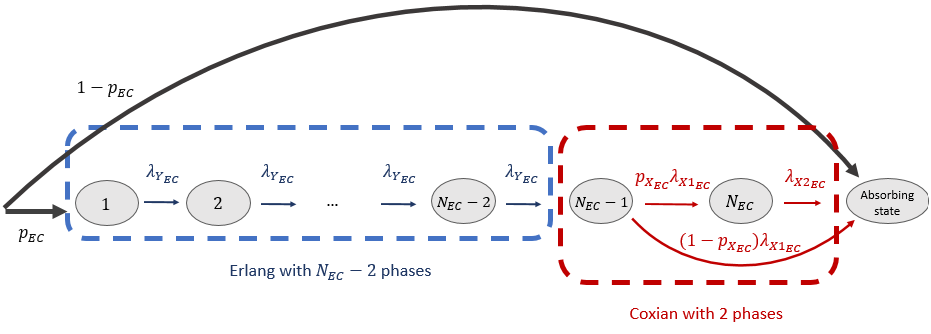}
\caption{An Erlang-Coxian (EC) distribution defined by six parameters $(p_{_{\text{EC}}}, \lambda_{Y_{\text{EC}}}, N_{_{\text{EC}}}, p_{X_{\text{EC}}}, \lambda_{X1_{\text{EC}}}, \lambda_{X2_{\text{EC}}})$.
     Each arrow represents an exponential transition time distribution.
     Calculations of these parameters depend on the normalized moments of the original distribution with an unknown shape \citep{Osogami2006}.}
\label{fig:simpleModel_transitionDiagram_EC} 
\end{figure}

Because of the truncation, the transition rate $\textbf{B}$ from $(1_+, n_-)$ to $(0, n_-)$ no longer represents a simple exponential transition time distribution.
In fact, the shape of this transition time distribution is often unknown but can be approximated to an Erlang-Coxian (EC) distribution.
As shown in Figure \ref{fig:simpleModel_transitionDiagram_EC}, a general EC distribution consists of exactly two Coxian phases and $N_{_{\text{EC}}}-2$ Erlang phases.
For a given distribution of unknown shape, \cite{Osogami2006} provided closed-form solutions to calculate the first three moments of the unknown distribution and the six parameters in the EC distribution that best matches the first three moments. 

\begin{figure*}[t]
\centering
\includegraphics[width=0.75\textwidth]{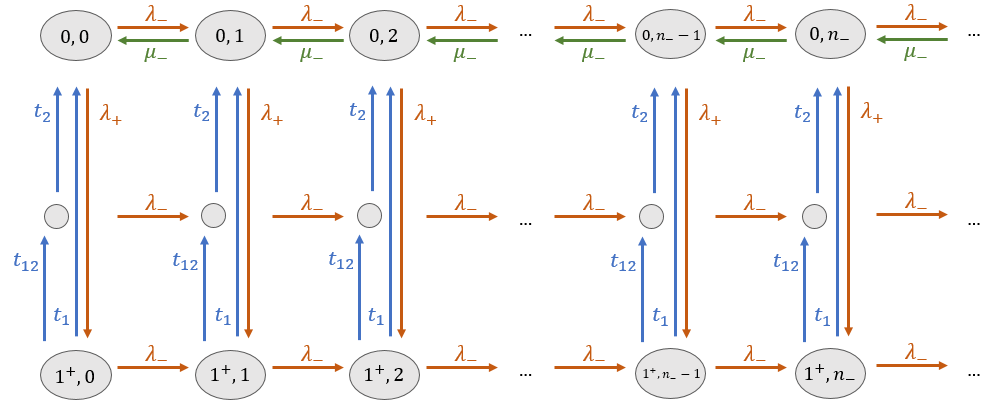}
\caption{The RDR-truncated, EC-approximated transition diagram for AI-negative, low-priority patient images in Model A assuming a with-CADt scenario, where the set of $t$'s correspond to the three Coxian rates (in red) in Figure \ref{fig:simpleModel_transitionDiagram_EC}.}
\label{fig:simpleModel_transitionDiagram_lowHBOts} 
\end{figure*}

When applying the EC-approximation method to the RDR-truncated transition diagram in Figure \ref{fig:simpleModel_transitionDiagram_lowHBO}, only the two-phase Coxian distribution is sufficient.
No Erlang phases are needed; hence, $p_{_{\text{EC}}}$, $N_{_{\text{EC}}}$, and $\lambda_{Y_{\text{EC}}}$ in Figure \ref{fig:simpleModel_transitionDiagram_EC} are 1, 2, and 0 respectively.
The non-exponential transition $\textbf{B}$ can then be explicitly expressed in terms of the approximated exponential transition rates $t$'s as shown in Figure \ref{fig:simpleModel_transitionDiagram_lowHBOts}, where
\begin{equation} \label{eqn:simpleModel_HBOts}
\begin{split}
t_1 & = (1-p_{X_{\text{EC}}})\lambda_{X1_{\text{EC}}}; \enspace \enspace \enspace
t_{12} = p_{X_{\text{EC}}}\lambda_{X1_{\text{EC}}}; \enspace \enspace \enspace
t_2 = \lambda_{X2_{\text{EC}}}. 
\end{split}
\end{equation}

Figure \ref{fig:simpleModel_transitionDiagram_lowHBOts} is a typical Markov chain transition diagram, and its transition rate matrix $M_{\text{A}}$ can be formed (see Section \ref{ECsec:queue_modelA_CADt} in Electronic Companions).
Using the matrix geometric method, an analysis method for quasi-birth–death processes where the Markov chain matrix has a repetitive block structure \citep{Stewart2009}, the state probability $p_{n_-}$ is computed.
Hence, the average waiting time per AI-negative, low-priority patient image, $W_{q_-}$, can be calculated;
\begin{equation}\label{eqn:Wq_low_CADt}
    W_{q_-} = \langle p_{n_-}\rangle/\lambda_- - 1/\mu_-.
\end{equation}

\subsection{Model B: Model A with emergent patient images}
\label{sec:queue_modelC}

Model B is similar to Model A but with the presence of emergent patient images ($f_{\text{em}} > 0$).
These emergent images are prioritized to the highest priority regardless of the presence of CADt devices.
Although the waiting time of the emergent subgroup can be studied, this work only focuses on the non-emergent, AI-positive, and AI-negative subgroups which are impacted by the CADt device.

\subsubsection{Model B in without-CADt scenario}
\label{sec:queue_modelC_noCADt}

In the standard of care without a CADt device, the presence of emergent class results in a two-priority-class queueing system: emergent and non-emergent classes.
For the emergent subgroup, $\mu_{\text{em}}$ denotes its radiologist's reading rate, and its arrival rate is given by
\begin{equation}\label{eqn:lambdaEM_fem}
    \lambda_\text{em} = f_{\text{em}}\lambda.
\end{equation}
The arrival rate for the non-emergent class is 
\begin{equation}\label{eqn:lambdaNonEM_fem}
    \lambda_\text{nonEm} = (1-f_{\text{em}})\lambda.
\end{equation}
Similar to Model A, because $\mu_D = \mu_{ND}$ and $N_{\text{rad}} = 1$, the effective reading rate for the non-emergent subgroup is $\mu_{\text{nonEm}} =\mu_D = \mu_{ND}$.

With only one radiologist on-site, the analysis of non-emergent, lower-priority class is exactly the same as that of the AI-negative class in Model A in the with-CADt scenario.
Figure \ref{fig:simpleModel_transitionDiagram_lowHBOts} (and Equation \ref{eqn:simpleModel_transitionMatrix_HBOts} in Electronic Companions) can be reused by replacing $\lambda_+$ with $\lambda_\text{em}$, $\lambda_-$ with $\lambda_\text{nonEm}$, $\mu_+$ with $\mu_{\text{em}}$, and $\mu_-$ with $\mu_{\text{nonEm}}$.
After solving for the state probability $p_{n_\text{nonEm}}$, the average waiting time per non-emergent patient image is given by Equation \ref{eqn:Wq_noCADt}.

\subsubsection{Model B in with-CADt scenario}
\label{sec:queue_modelC_CADt}

When a CADt is included in the workflow, three priority classes exist: emergent (highest priority), AI-positive (middle priority), and AI-negative (lowest priority) classes.
With the presence of emergent patients, the arrival rates of AI-positive and AI-negative classes are now
\begin{equation}\label{eqn:lambdaPlus_fem}
    \lambda_{+} = \big[\pi\text{Se} + (1-\pi)(1-\text{Sp})\big ](1-f_{\text{em}})\lambda, \enspace \text{and}
\end{equation}
\begin{equation}\label{eqn:lambdaMinus_fem}
    \lambda_{-} = \big[\pi(1-\text{Se}) + (1-\pi)\text{Sp}\big](1-f_{\text{em}})\lambda.
\end{equation}
Their reading rates are given by Equations \ref{eqn:muPlus_eff} and \ref{eqn:muMinus_eff}.
However, because $\mu_D = \mu_{ND}$, the reading rates for the AI-positive and AI-negative subgroups are the same; $\mu_+ = \mu_- = \mu_D = \mu_{ND}$.
Similar to Model A in with-CADt scenario, we apply the RDR method and solve for the AI-positive and AI-negative systems separately.

For the AI-positive subgroup, it is noted that an AI-positive patient image can only be interrupted by emergent patient images and will not be impacted by any AI-negative patient images.
Therefore, the emergent and AI-positive subgroups form a two-priority-class queueing system which can be solved using the framework developed for the non-emergent subgroup in the without-CADt scenario.
Figure \ref{fig:simpleModel_transitionDiagram_lowHBOts} (and Equation \ref{eqn:simpleModel_transitionMatrix_HBOts} in Electronic Companions) can be reused by replacing $\lambda_+$ with $\lambda_\text{em}$, $\lambda_-$ with $\lambda_+$, $\mu_+$ with $\mu_{\text{em}}$, and $\mu_-$ with $\mu_+$.
The state probability $p_{n_+}$ for the AI-positive subgroup is calculated, from which the average waiting time per AI-positive patient image is given by Equation \ref{eqn:Wq_high_CADt}.

The calculation for the AI-negative, lowest-priority subgroup involves states $(n_{\text{em}}, n_+, n_-)$ defined by the number of emergent, AI-positive, and AI-negative patient images in the system.
An AI-negative patient image can be interrupted by either an emergent or an AI-positive patient image.
The arrival time of the interrupting case denotes the start of a busy period, which is defined as the time period during which a radiologist is too busy for AI-negative cases.
While the radiologist is reading the interrupting case, new emergent and/or AI positive images may enter the system, which further delays the review of the interrupted AI-negative case.
Once all the higher-priority images are reviewed, the radiologist then resumes the reading of the interrupted AI-negative patient image, and the busy period ends.

\begin{figure*}[t]
\centering
\includegraphics[width=0.75\textwidth]{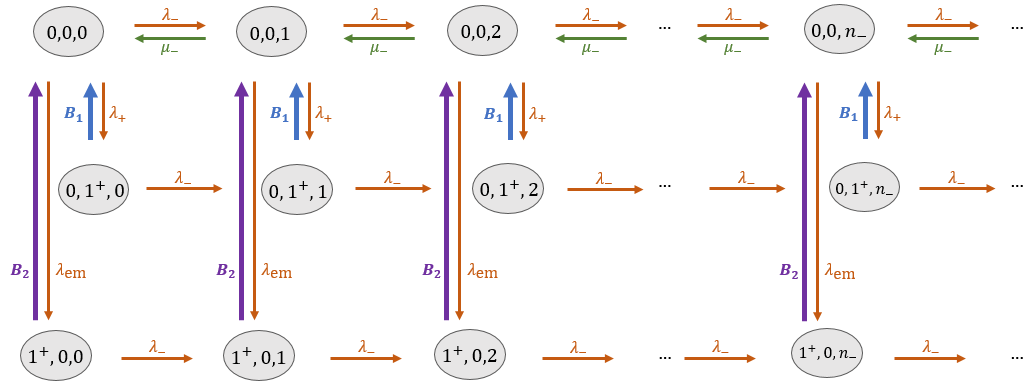}
\caption{The RDR-truncated transition diagram for AI-negative class in Model B in the with-CADt scenario.
The state $(n_{\text{em}}, n_+, n_-)$ is defined by the numbers of emergent, AI-positive, AI-negative patient images.
Thick arrows represent the non-exponential transition rates \textbf{$B_1$} and \textbf{$B_2$} of the two busy periods.}
\label{fig:modelC_transitionDiagram_lowHBO} 
\end{figure*}

Due to the different arrival and reading rates between the emergent and AI-positive patient images, the dependence of AI-negative busy period on the two subgroups are different.
As \cite{HarcholBalter2005} discussed, one must keep track of the state at which the busy period starts and the state at which the busy period ends.
With only one radiologist, Model B has only two distinct busy periods:
\begin{itemize}
    \item \textbf{$B_1$}: (0, $1^+$, $n_-$) $\rightarrow$ (0, 0, $n_-$)
    \item \textbf{$B_2$}: ($1^+$, 0, $n_-$) $\rightarrow$ (0, 0, $n_-$)
\end{itemize}
Here, $B_1$ and $B_2$ are the rates of the two busy periods and are explicitly shown as two non-exponential transitions in Figure \ref{fig:modelC_transitionDiagram_lowHBO}.

Just like the AI-negative system in Model A, one must first calculate the first three moments for each busy period and approximate each distribution using a two-phase Coxian distribution.
With three priority classes and two busy periods, the approximation involves the inter-level passage times from the AI-positive transition diagram, from which a transition probability matrix as well as the transition rate matrix are determined (see $M_{\text{B}}$ in Section \ref{ECsec:queue_modelC_CADt}).
From transition rate matrix, the state probability $p_{n_-}$ can be solved via conventional matrix geometric method.
Once $p_{n_-}$ is determined, the average waiting time per AI-negative patient image $W_{q_-}$ can be calculated via Equation \ref{eqn:Wq_low_CADt}.

\subsection{Model C: Model B with two radiologists}
\label{sec:queue_modelD}

Model C extends Model B by adding one extra radiologist on-site $N_{\text{rad}} = 2$.
The arrival rates for the emergent, non-emergent, AI-positive, and AI-negative classes remain the same (Equations \ref{eqn:lambdaEM_fem} - \ref{eqn:lambdaMinus_fem}).
Because $\mu_D = \mu_{ND}$, the reading rates for the non-emergent, AI-positive, and AI-negative subgroups are the same; $\mu_+ = \mu_- = \mu_{\text{nonEm}}$.
Because of the extra radiologist, the traffic intensity $\rho$ has a factor of two; $\rho=\lambda/2\mu$.
It should be noted that Model C has the same settings as the example in \cite{HarcholBalter2005}.

\subsubsection{Model C in without-CADt scenario}
\label{sec:queue_modelD_noCADt}

\begin{figure*}[t]
\centering
\includegraphics[width=0.8\textwidth]{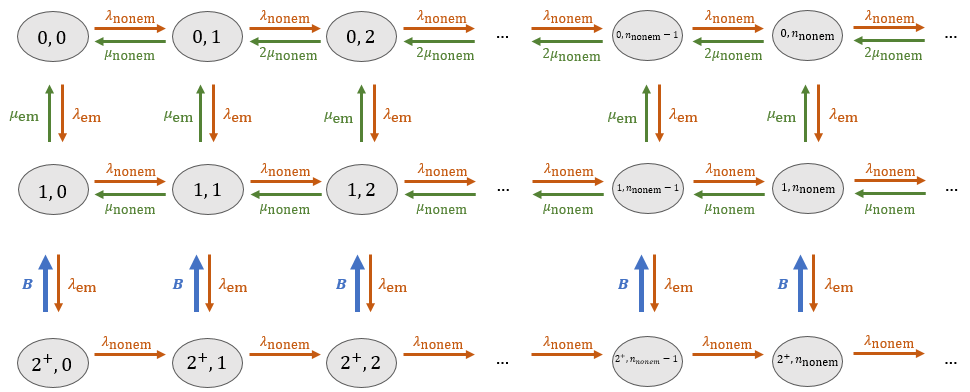}
\caption{The RDR-truncated transition diagram for non-emergent patient images in Model C in the without-CADt scenario.
The states $(n_{\text{em}}, n_{\text{nonEm}})$ are defined by the numbers of emergent and non-emergent cases in the system.}
\label{fig:modelD_transitionDiagram_noCADt} 
\end{figure*}

With no CADt devices, the RDR-truncated transition diagram for the non-emergent, lower-priority class is given by Figure \ref{fig:modelD_transitionDiagram_noCADt}.
Given two radiologists on-site, a non-emergent image can depart the system only when $n_{\text{em}} < 2$, and hence the truncation of states starts when $n_{\text{em}} = 2$.
Moreover, when $n_{\text{em}} = 0$, both radiologists are available for non-emergent patient images.
Thus, the first row has a leaving rate $2\mu_{\text{nonEm}}$, except the transition from $(0, 1)$ to $(0, 0)$ when only one radiologist has work to do.
When $n_{\text{em}} = 1$ (the second row), only one of the two radiologists is available to review a non-emergent case, resulting in a leaving rate of $1\mu_{\text{nonEm}}$.
When $n_{\text{em}} \geq 2$, both radiologists are busy handling emergent cases.
Since no radiologist is available for non-emergent images, their leaving rate is $0$, and no non-emergent images can leave the system.
To approximate the transition rate $\textbf{B}$ in Figure \ref{fig:modelD_transitionDiagram_noCADt}, the same two-phase Coxian approximation described in Models A and B is applied.

The transition rate matrix $M_{\text{C}_{\text{noCADt}}}$ for Figure \ref{fig:modelD_transitionDiagram_noCADt} can be found in Section \ref{ECsec:queue_modelD_noCADt}.
From $M_{\text{C}_{\text{noCADt}}}$, the state probability $p_{n_\text{nonEm}}$ is determined, and the average waiting time per non-emergent patient image is given by Equation \ref{eqn:Wq_noCADt}.

\subsubsection{Model C in with-CADt scenario}
\label{sec:queue_modelD_CADt}

In the with-CADt scenario, the calculations for AI-positive (middle-priority) and AI-negative (lowest-priority) subgroups are separated.

The queueing system for the AI-positive subgroup consists of two priority classes: the emergent and AI-positive classes, and the framework developed for the non-emergent subgroup in the without-CADt scenario can be reused.
By replacing $\lambda_{\text{nonEm}}$ with $\lambda_+$ and $\mu_{\text{nonEm}}$ with $\mu_+$ in Figure \ref{fig:modelD_transitionDiagram_noCADt} and Equation \ref{eqn:modelD_transitionMatrix_noCADt}, the state probability for the AI-positive subgroup $p_{n_+}$ can be computed.
And the average waiting time per AI-positive patient image $W_{q_+}$ is given by Equation \ref{eqn:Wq_high_CADt}.

\begin{figure*}[t]
\centering
\includegraphics[width=0.85\textwidth]{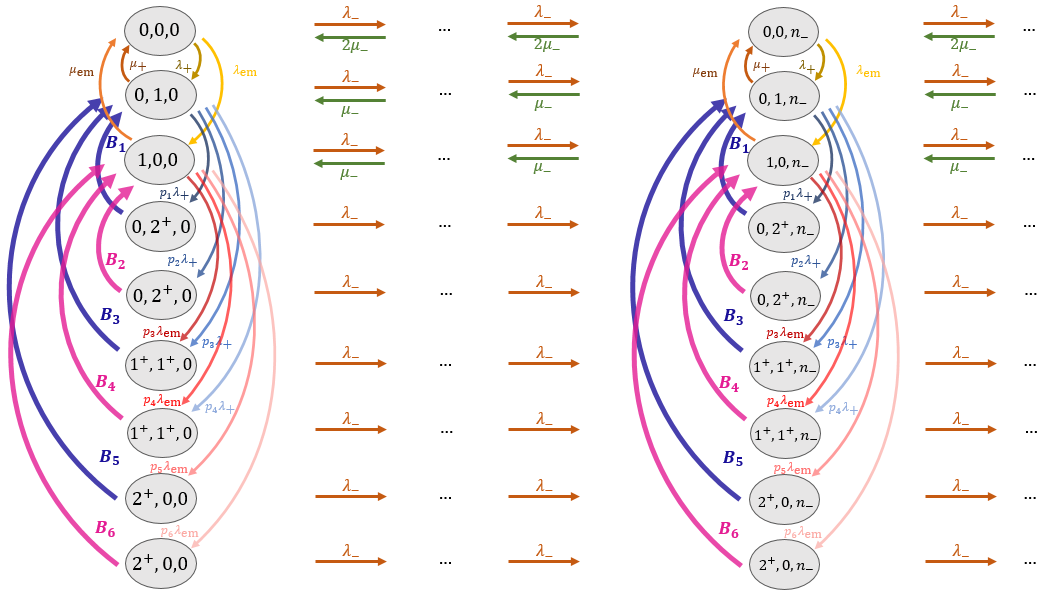}
\caption{The RDR-truncated transition diagram for AI-negative patient images in Model C in the with-CADt scenario.
The state is defined as $(n_{\text{em}}, n_+, n_-)$, and states with $n_{\text{em}} + n_+ \geq 2$ are truncated.
A total of 6 busy periods are identified.
Each busy period $i$ has a transition rate $\textbf{B}_i$ along with a probability that it ends at a certain state given that it starts with a particular state.}
\label{fig:modelD_transitionDiagram_lowHBO} 
\end{figure*}

The approach to analyze the AI-negative, lowest-priority subgroup is similar to the analysis of the AI-negative cases in Model B.
Recall that a state is defined as $(n_{\text{em}}, n_+, n_-)$ and that a busy period is defined by the time duration in which all the radiologists on-site are too busy for AI-negative patient images.
With two radiologists, a busy period may start from one of the three situations: when there are two emergent cases, when there are one emergent and one AI-positive case, or when there are two AI-positive cases.
On the other hand, the busy period ends when one radiologist is handling either an emergent case or an AI-positive case such that the other radiologist is available for the AI-negative case.
Therefore, instead of two busy periods in Model B, adding one extra radiologist increases the total number of busy periods to six:
\begin{itemize}
    \item \textbf{$B_1$}: (0, $2^+$, $n_-$) $\rightarrow$ (0, $1^+$, $n_-$)
    \item \textbf{$B_2$}: (0, $2^+$, $n_-$) $\rightarrow$ ($1^+$, 0, $n_-$)
    \item \textbf{$B_3$}: ($1^+$, $1^+$, $n_-$) $\rightarrow$ (0, $1^+$, $n_-$)
    \item \textbf{$B_4$}: ($1^+$, $1^+$, $n_-$) $\rightarrow$ ($1^+$, 0, $n_-$)
    \item \textbf{$B_5$}: ($2^+$, 0, $n_-$) $\rightarrow$ (0, $1^+$, $n_-$)
    \item \textbf{$B_6$}: ($2^+$, 0, $n_-$) $\rightarrow$ ($1^+$, 0, $n_-$)
\end{itemize}

Figure \ref{fig:modelD_transitionDiagram_lowHBO} shows the RDR-truncated transition diagram for AI-negative subgroup.
Note that states $(0, 2^+, n_-)$, $(1^+, 1^+, n_-)$, and $(2^+, 0, n_-)$ are duplicated because their corresponding arrival rates also depends on the probabilities that the busy period ends at a particular state i.e. either $(0, 1^+, n_-)$ or $(1^+, 0, n_-)$. 
For example, $p_1$ denotes the conditional probability that the busy period ends at $(0, 1^+, n_-)$ given that it starts at $(0, 2^+, n_-)$.

Before solving for Figure \ref{fig:modelD_transitionDiagram_lowHBO}, one must compute the conditional probability and the first three moments of each busy period, from which the transition rates can be approximated.
The calculation is discussed in Section \ref{ECsec:queue_modelD_CADt}, where the AI-positive transition diagram for inter-level passage times is presented, and the transition probability matrix is constructed.

\begin{figure}[t]
\centering
\includegraphics[width=\columnwidth]{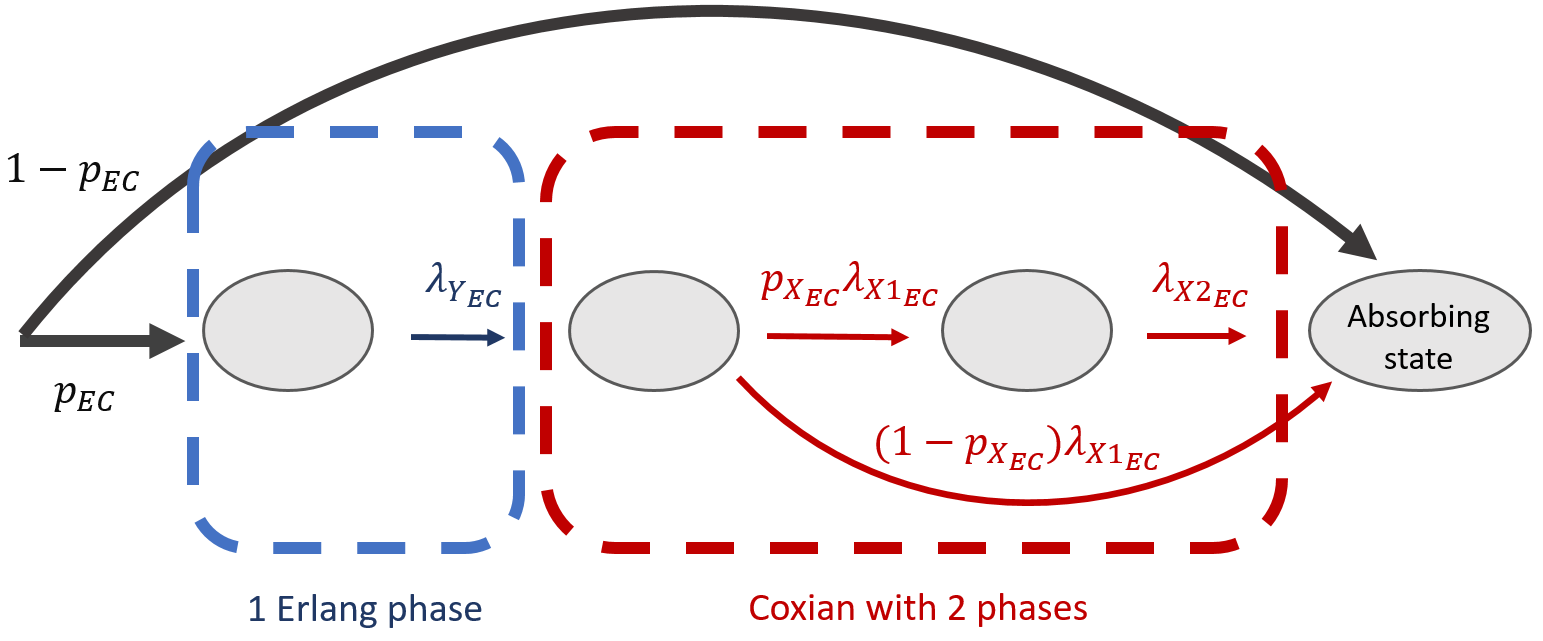}
\caption{An Erlang-Coxian (EC) distribution with one Erlang phase and two Coxian phases.
See \cite{Osogami2006} for the closed form solutions to calculate these parameters.}
\label{fig:modelD_transitionDiagram_ECwith1Erlang} 
\end{figure}

Each busy period is approximated using the EC distribution (Figure \ref{fig:simpleModel_transitionDiagram_EC}).
However, unlike Model B in which two-phase Coxian is sufficient for all busy periods, $B_2$ and $B_5$ in Model C require an extra Erlang phase, as shown in Figure \ref{fig:modelD_transitionDiagram_ECwith1Erlang}.
With an extra phase, two extra parameters $t_0$ and $t_{01}$ are needed to approximate $B_2$ and $B_5$.
\begin{equation} \label{eqn:modelD_HBOts}
\begin{split}
t_0 & = (1-p_{_{\text{EC}}})\lambda_{Y_{\text{EC}}}; \enspace \enspace \enspace \enspace \enspace
t_{01} = p_{_{\text{EC}}}\lambda_{Y_{\text{EC}}}; \\
t_1 & = (1-p_{X_{\text{EC}}})\lambda_{X1_{\text{EC}}}; \enspace \enspace \enspace 
t_{12} = p_{X_{\text{EC}}}\lambda_{X1_{\text{EC}}}; \enspace \enspace \enspace
t_2 = \lambda_{X2}. 
\end{split}
\end{equation}

Once all six busy periods are approximated, the transition rate matrix for the AI-negative, lowest-priority class can be constructed from Figure \ref{fig:modelD_transitionDiagram_lowHBO}. (See Section \ref{ECsec:queue_modelD_CADt}.)
Like before, the corresponding state probability $p_{n_-}$ can be solved by the matrix geometric method.
And, the average waiting time per AI-negative patient image $W_{q_-}$ can be calculated via Equation \ref{eqn:Wq_low_CADt}.

For $N_{\text{rad}} \geq 3$, the same approach can be applied.
However, as the number of busy periods increases, the transition rate matrix will grow in size drastically, especially when more Erlang phases are required for the busy period approximation.

\subsection{Model D: Model B with different reading rates}
\label{sec:queue_modelE}
Model D extends Model B by differentiating the radiologist's reading rate between the diseased and non-diseased subgroups ($N_{\text{rad}} = 1$, $f_{\text{em}} > 0$, and $\mu_D \neq \mu_{ND}$).
The arrival rates for the emergent, non-emergent, AI-positive, and AI-negative classes remain the same (Equations \ref{eqn:lambdaEM_fem} - \ref{eqn:lambdaMinus_fem}).
However, because $\mu_D \neq \mu_{ND}$, the reading rates for non-emergent, AI-positive, and AI-negative subgroups depend on disease prevalence $\pi$, positive predictive value $\text{PPV}$, and negative predictive value $\text{NPV}$ (Equations \ref{eqn:muNonEm}-\ref{eqn:muMinus_eff}).

\subsubsection{Model D in without-CADt scenario}
\label{sec:queue_modelE_noCADt}

\begin{figure*}[t]
\centering
\includegraphics[width=0.85\textwidth]{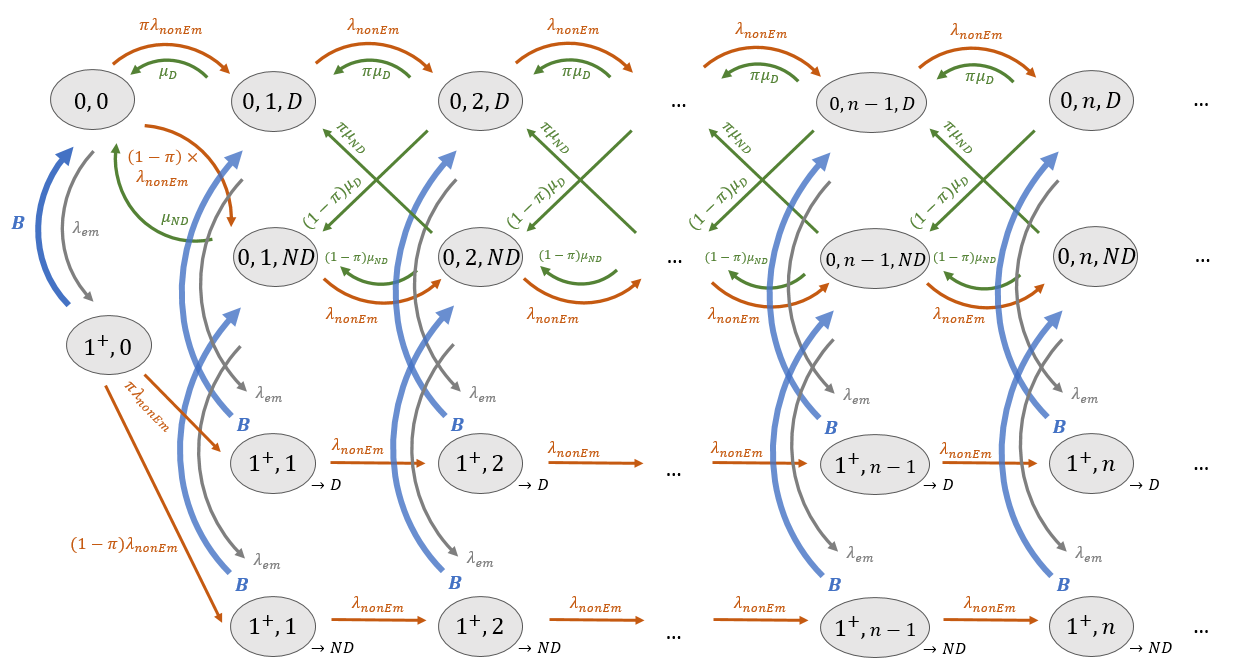}
\caption{The RDR-truncated transition diagram for non-emergent patient images in Model D in a without-CADt scenario.
The state is defined by the number of emergent patient images (either 0 or 1$^+$), number of non-emergent patient images $n$, and the disease status of case that the one radiologist is reviewing (either $D$ for diseased or $ND$ for non-diseased).
The ``$\rightarrow D$" and ``$\rightarrow ND$" in the truncated states keep track of the disease status of the interrupted, lower-priority case.}
\label{fig:modelE_transitionDiagram_noCADt} 
\end{figure*}  

The without-CADt scenario has two priority classes: emergent and non-emergent patient images.
Within the non-emergent class, two groups of patient images (diseased and non-diseased) are reviewed in a first-in-first-out (FIFO) basis.
The corresponding transition diagram is shown in Figure \ref{fig:modelE_transitionDiagram_noCADt}.
As usual, the state keeps track of $n_{\text{em}}$ and $n_{\text{nonEm}}$.
In addition, because of the different reading rates between the diseased and non-diseased subgroups, the state must also keep track of the disease status of the image that the radiologist is reviewing.
Therefore, the state is defined as $(n_{\text{em}}, n_{\text{nonEm}}, i)$, where $i$ is either $D$ (i.e. the radiologist is working on a diseased image) or $ND$ (i.e. the radiologist is working on a non-diseased image).
Furthermore, one must pay attention to how the busy period starts and ends.
For example, if the radiologist reading a diseased image is interrupted by the arrival of an emergent image i.e. $(0, n, D) \rightarrow (1^+, n, D)$, the state must go back to $(0, n, D)$ and not to $(0, n, ND)$ when the busy period is over.
This property is guaranteed by having two sets of truncated states: $(1^+, n)_{\rightarrow D}$ that can only interact with $(0, n, D)$ and $(1^+, n)_{\rightarrow ND}$ that can only interact with $(0, n, ND)$.

The corresponding transition rate matrix of Figure \ref{fig:modelE_transitionDiagram_noCADt} is given in Section \ref{ECsec:queue_modelE_noCADt}. 
Note that, although Figure \ref{fig:modelE_transitionDiagram_noCADt} has two busy periods per column (one for ``$\rightarrow D$'' and the other for ``$\rightarrow ND$''), they both describe the same transition time when at least one emergent image is in the system.
Therefore, only one unique set of $t$-parameters is calculated to approximate both busy periods.

\subsubsection{Model D in with-CADt scenario}
\label{sec:queue_modelE_CADt}

The calculation for AI-positive (middle-priority) and AI-negative (lowest-priority) subgroups are separated. 

Because AI-positive patient images are not impacted by AI-negative cases, the emergent and AI-positive subgroups form a two-priority-class queueing system.
The transition rate matrix $M_{D_{\text{noCADt}}}$ from Figure \ref{fig:modelE_transitionDiagram_noCADt} can be reused to analyze the queueing of AI-positive patient images.
By replacing $\lambda_{\text{nonEm}}$ by $\lambda_+$ (Equation \ref{eqn:lambdaPlus_fem}), $\mu_{\text{nonEm}}$ by $\mu_+$ (Equation \ref{eqn:muPlus_eff}), and $\pi$ by $\text{PPV}$, the state probability for the AI-positive subgroup $p_{n_+}$ is calculated via standard matrix geometric method.
The average waiting time per AI-positive patient image $W_{q_+}$ is then given by Equation \ref{eqn:Wq_high_CADt}.

\begin{figure*}[t]
\centering
\includegraphics[width=0.85\textwidth]{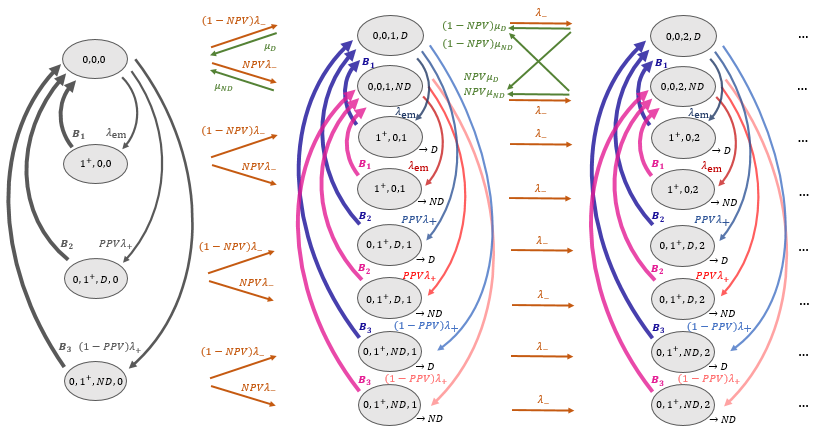}
\caption{The RDR-truncated transition diagram for AI-negative subgroup in Model E in the with-CADt scenario.
State is defined as $(n_{\text{em}}, n_+, i, n_-, j)$, where $i$ (or $j$) indicate the disease status of the AI-positive (or AI-negative) case the radiologist is reading.
}
\label{fig:modelE_transitionDiagram_lowHBO}  
\end{figure*}  

For the AI-negative, lowest-priority class, the full definition of state $(n_{\text{em}}, n_+, i, n_-, j)$.
$i$ is either $D$ or $ND$, indicating whether the radiologist is working on a diseased, AI-positive case or a non-diseased, AI-positive case respectively.
The disease status of an AI-negative case that the radiologist is reading is represented by the $j$ which is either $D$ or $ND$.
Because we only have one radiologist, $i$ and $j$ cannot appear simultaneously; the one radiologist can only handle an AI-positive or AI-negative case but not both at the same time.

Figure \ref{fig:modelE_transitionDiagram_lowHBO} shows the RDR-truncated transition diagram for the AI-negative subgroup.
There are three unique busy periods with the corresponding transition rates $B_i$:
\begin{itemize}
    \item \textbf{$B_1$}: ($1^+$, 0, $n_-$) $\rightarrow$ (0, 0, $n_-$, $j$)
    \item \textbf{$B_2$}: (0, $1^+$, $D$, $n_-$) $\rightarrow$ (0, 0, $n_-$, $j$)
    \item \textbf{$B_3$}: (0, $1^+$, $ND$, $n_-$) $\rightarrow$ (0, 0, $n_-$, $j$)
\end{itemize}
For each busy period, the truncated state is duplicated with either ``$\rightarrow D$" or ``$\rightarrow ND$" such that the system can return to the state with the correct disease status $j$ when the busy period is over.

Like before, for each unique busy period, its conditional probability and first three moments are determined from the transition probability matrix (see Section \ref{ECsec:queue_modelE_CADt}).
And, each unique busy period has a set of $t$-parameters (Equation \ref{eqn:simpleModel_HBOts}) approximated from a two-phase Coxian distribution.
With the approximated busy period transitions, a transition rate matrix can be constructed for the AI-negative subgroup from Figure \ref{fig:modelE_transitionDiagram_lowHBO} (see Section \ref{ECsec:queue_modelE_CADt}).
The state probability $p_{n_-}$ is then solved, and the average waiting time per AI-negative patient image $W_{q_-}$ can be calculated via Equation \ref{eqn:Wq_low_CADt}.

\section{Simulation}
\label{sec:simulation}

To verify the analytical results from our theoretical queueing approach, a Monte Carlo software was developed using Python to simulate the flow of patient images in a clinic with and without a CADt device.
A workflow model is defined by a set of input parameters \{$f_{\text{em}}$, $\pi$, $\rho$, $\mu$, $N_{\text{rad}}$, $\text{Se}$, and $\text{Sp}$\}.

During the simulation, a new patient image entry is randomly generated with a timestamp that follows a Poisson distribution at an overall arrival rate of $\lambda$, which is computed from the user-inputs (traffic $\rho$ and radiologist's reading rates $\mu$).
Each patient image is randomly assigned with an emergency status (emergent or non-emergent) based on the input emergency fraction $f_{\text{em}}$.
If the patient image is emergent, a reading time is randomly generated from an exponential distribution with a reading rate of $\mu_{\text{em}}$.
If the patient image is non-emergent, a disease status (diseased or non-diseased) is randomly assigned based on the input disease prevalence $\pi$.
The reading time for this non-emergent patient image is also randomly drawn from an exponential distribution with a reading rate of either $\mu_D$ if it is diseased or $\mu_{ND}$ if it is non-diseased.
Each non-emergent patient image is also assigned with an AI-call status (positive or negative) based on its disease status and the input AI accuracy ($\text{Se}$ and $\text{Sp}$).
The patient image is then simultaneously placed into two worlds: one with a CADt device and one without.

In a without-CADt world, the incoming patient image is either a higher-priority case (if it is emergent) or a lower-priority case (if non-emergent).
If the patient image is emergent, the case is prioritized over all non-emergent patient images in the system and is placed at the end of the emergent-only queue.
Otherwise, the patient image is non-emergent and is placed at the end of the current reading queue.
In time, when its turn comes, this patient image is read by one of the radiologists and is then removed from the queue.
Two pieces of information are recorded for this simulated patient image.
One is its waiting time defined as the difference between the time when the image enters the queue and when it leaves the queue.
In addition, the number of emergent and non-emergent patient images in the queue right before the arrival of the new patient image are also recorded to study the state probability distribution.

Alternatively, this very same patient image is placed in the with-CADt world.
This image has either a high priority (if emergent), a middle priority (if AI-positive), or a low priority (if AI-negative).
If the patient image is emergent, the case is prioritized over all AI-positive and AI-negative patient images in the system and is placed at the end of the emergent-only queue.
If the patient image is AI-positive, the case is prioritized over all AI-negative images and is placed at the end of the queue consisting of only emergent and AI-positive patient images.
Otherwise, the patient image is AI-negative and is placed at the end of the current reading queue.
The reading time for this patient image in the with-CADt world is identical to its reading time in the without-CADt world.
However, due to the re-ordering by the CADt device, its waiting time in the with-CADt world may be different from that in the without-CADt world.
For every patient image, the difference between the two waiting times in the two worlds can be calculated to determine whether the use of the CADt device results in a time-saving or time delay for this image.
In addition to its waiting time, the number of emergent, AI-positive, and AI-negative patient images right before the arrival of the new patient image are also recorded.

To simulate a big enough sample size, a full simulation includes 200 simulations, each of which contains roughly 2,000 patients.
From all simulations, the waiting times from all diseased patient images are histogrammed from which the mean value and the 95\% confidence intervals are determined.

\section{Time-saving effectiveness evaluation metric}
\label{sec:metric}

We define a metric to quantitatively assess the time-saving effectiveness of a given CADt device.
Both theoretical and simulation approaches output the mean waiting time per diseased patient image $W_{\text{D}}$ in both with- and without-CADt scenarios.

Without a CADt device, since the arrival process is random, the average waiting time per non-emergent patient image $W^{\text{no-CADt}}_{\text{nonEm}}$ is the same as $W^{\text{no-CADt}}_{\text{D}}$ i.e. 
\begin{equation}\label{eqn:WqD_noCADt}
    W^{\text{no-CADt}}_{\text{D}} = W^{\text{no-CADt}}_{\text{nonEm}} = W_{q_{\text{nonEm}}}.
\end{equation}
When a CADt device is included in the workflow, the average waiting time per diseased and non-diseased patient images are no longer the same because the diseased images are more likely to be prioritized by the CADt. 
To calculate $W^{\text{CADt}}_{\text{D}}$, we first compute the average waiting time per AI-positive ($W^{\text{CADt}}_{+} = W_{q_+}$) and per AI-negative ($W^{\text{CADt}}_{-} = W_{q_-}$) patient image based on the mathematical frameworks discussed in Section \ref{sec:models}.
By definition, the average waiting time for the diseased subgroup $W^{\text{CADt}}_{\text{D}}$ is 
\begin{equation*}
    W^{\text{CADt}}_{\text{D}} \equiv \frac{\text{Total waiting time from all diseased patient images}}{\text{Number of diseased patient images}}.
\end{equation*}
Note that the total waiting time from all diseased patients is the sum of the total waiting time from the true-positive (TP) subgroup and that from the false-negative (FN) subgroup.
Let $N_{\text{TP}}$, $N_{\text{FN}}$, and $N_{\text{D}}$ be the number of TP patient images, that of FN patient images, and that of diseased images.
$W^{\text{CADt}}_{\text{D}}$ can be rewritten as
\begin{equation*}
    W^{\text{CADt}}_{\text{D}} = \frac{W^{\text{CADt}}_{+} \times N_{\text{TP}} + W^{\text{CADt}}_{-} \times N_{\text{FN}}}{N_{\text{D}}}.
\end{equation*}
Because $N_{\text{TP}}/N_{\text{D}}$ and $N_{\text{FN}}/N_{\text{D}}$ are, by definition, $\text{Se}$ and $1-\text{Se}$, we have
\begin{equation}\label{eqn:WqD_CADt}
    W^{\text{CADt}}_{\text{D}} = W^{\text{CADt}}_{+} \times \text{Se} + W^{\text{CADt}}_{-} \times (1-\text{Se}).
\end{equation}    

To quantify the time-saving effectiveness of a CADt device for diseased patient images, we define a time performance metric $\delta W_\text{D}$ as the difference in mean waiting time per diseased image in the with-CADt and that in the without-CADt scenario:
\begin{equation}\label{eqn:dWqD}
    \delta W_{\text{D}} \equiv W^{\text{CADt}}_{\text{D}} - W^{\text{no-CADt}}_{\text{D}}.
\end{equation}
It should be noted that, besides the explicit dependence on AI sensitivity in Equation \ref{eqn:WqD_CADt}, $\delta W_{\text{D}}$ also depends on AI specificity and all the clinical factors in the calculation of $W^{\text{CADt}}_{+}$, $W^{\text{CADt}}_{-}$, and $W^{\text{no-CADt}}_{\text{nonEm}}$.

Based on its definition, a negative $\delta W_{\text{D}}$ implies that, on average, a diseased patient image is reviewed earlier when the CADt device is included in the workflow than when it is not.
The more negative $\delta W_{\text{D}}$ is, the more time is saved, and the more effective the CADt device is.
If $\delta W_{\text{D}} = 0$, the presence of CADt device does not bring any benefit for the diseased patient images.
If $\delta W_{\text{D}}$ is positive, the review of a diseased patient image is delayed on average, and the CADt device brings more risks than benefits to the diseased subgroup.

It should also be noted that the amount of time savings for other subgroups can be defined similarly.
For example, for the non-diseased subgroup, the average waiting time per non-diseased patient image in the without-CADt scenario, $W^{\text{no-CADt}}_{\text{ND}}$, is 
\begin{equation*}
    W^{\text{no-CADt}}_{\text{ND}} = W^{\text{no-CADt}}_{\text{nonEm}}.
\end{equation*}
When the CADt device is included in the workflow, the average waiting time per non-diseased patient image, $W^{\text{CADt}}_{\text{ND}}$, becomes 
\begin{equation*}
    W^{\text{CADt}}_{\text{ND}} = W^{\text{CADt}}_{+} \times (1-\text{Sp}) + W^{\text{CADt}}_{-} \times \text{Sp},
\end{equation*}
where the first and second terms correspond to the false-positive and true-negative subgroups respectively.
$\delta W_{\text{ND}}$ can then be defined to describe the average wait-time difference between the with-CADt and without-CADt scenarios for the non-diseased subgroup.

\section{Results and Discussion}
\label{sec:results}

\begin{figure}[t]
\centering
\includegraphics[width=\columnwidth]{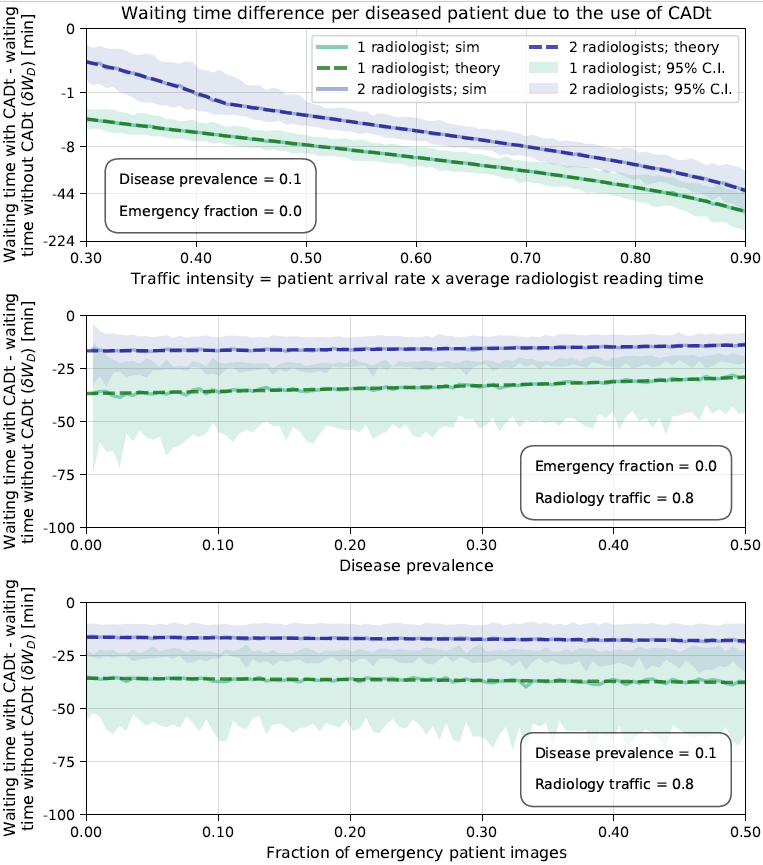}
\caption{Amount of time saved per diseased patient image as a function of (top) traffic intensity, (middle) disease prevalence, and (bottom) emergency fraction.
Green and blue lines represent scenarios with one and two radiologists respectively.
Dashed lines are theoretical $\delta W_D$, and the solid lines represent the mean time-saving effectiveness from simulation.
Shaded areas are the 95\% confidence intervals (C.I.s) from simulation.
The average reading time for an emergent image is set at 5 minutes, whereas the average reading time for the diseased and non-diseased subgroups are both 10 minutes.
}
\label{fig:SPIE_1D} 
\end{figure}

Top plot in Figure \ref{fig:SPIE_1D} shows the time saved per diseased patient images as a function of traffic intensity $\rho$ for one and two radiologists on-site without any emergent patient images.
Assuming a disease prevalence $\pi$ of 10\%, an AI sensitivity of 95\%, a specificity of 89\%, an average image reading time of 10 minutes for both diseased and non-diseased subgroups, and one radiologist on-site, the time saving is significantly improved from about 2 minutes in a quiet, low-volume clinic (radiology traffic intensity of 0.3) to about an hour in a relatively busy clinic (radiology traffic intensity of 0.8).
At a traffic intensity $\rho$ of 0.8, the impact due to disease prevalence is found to be small (see middle plot in Figure \ref{fig:SPIE_1D}).
Overall, the time-saving effectiveness of the device is also found to be more evident with only one radiologist on-site compared to two.
Bottom plot in Figure \ref{fig:SPIE_1D} shows the impact on the time-saving effectiveness due to the presence of emergent patient images with the highest priority that overrides any AI prioritization.
The amount of time saved per diseased image without any emergent patients ($f_{\text{em}} = 0$) is more-or-less the same as that with $f_{\text{em}} = 50$\%.
This is likely because the amount of delay caused by emergent patient images in a without-CADt scenario is similar to that in the with-CADt scenario.

\begin{figure}[t]
\centering
\includegraphics[width=\columnwidth]{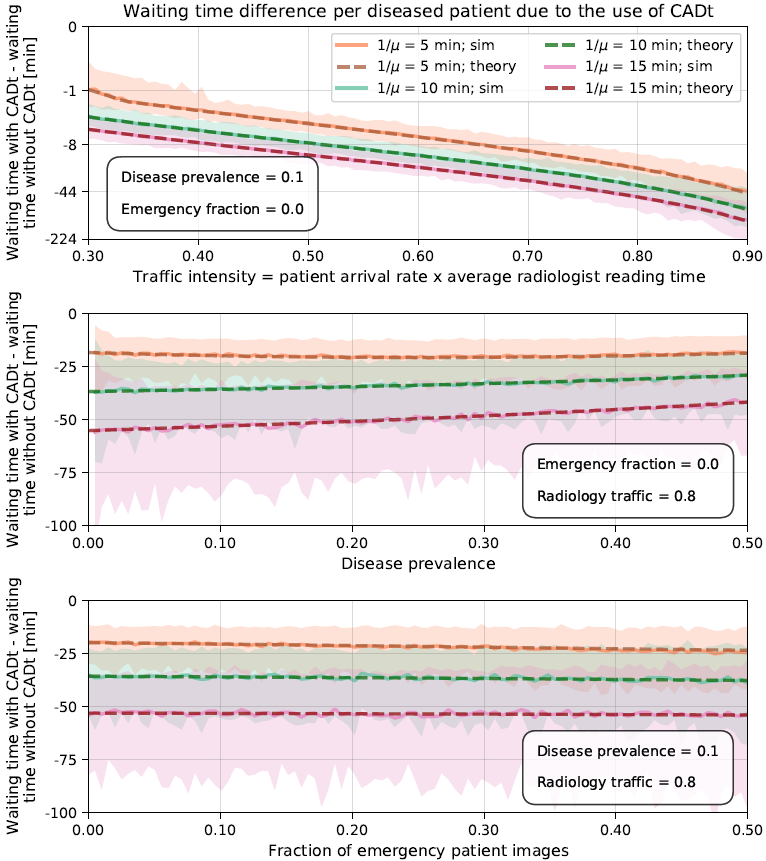}
\caption{Amount of time saved per diseased patient as a function of (top) traffic intensity, (middle) disease prevalence, and (bottom) emergency fraction.
Only one radiologist is on-site, and its average reading times for emergent and diseased patient images are set at 5 minutes and 10 minutes respectively.
The average reading time for non-diseased patient images varies between 5 minutes (orange), 10 minutes (green), and 15 minutes (red).
Dashed lines are theoretical $\delta W_D$, and the solid lines represent the mean time-saving effectiveness from simulation.
Shaded areas are the 95\% confidence intervals (C.I.s) from simulation.
Note that the green set of lines here is identical to that in Figure \ref{fig:SPIE_1D}.
}
\label{fig:deltaWD_clinicalFactors} 
\end{figure}  

The effect of having different radiologist's reading rates for diseased and non-diseased subgroups are shown in Figure \ref{fig:deltaWD_clinicalFactors}.
The overall dependence on traffic intensity, disease prevalence, and emergency fraction is similar to that in Figure \ref{fig:SPIE_1D}.
However, more time is saved for diseased patient images when $\mu_D < \mu_{ND}$ i.e. when a radiologist takes more time on average to read a non-diseased image than a diseased image.

\begin{figure*}[t]
\centering
\includegraphics[width=\textwidth]{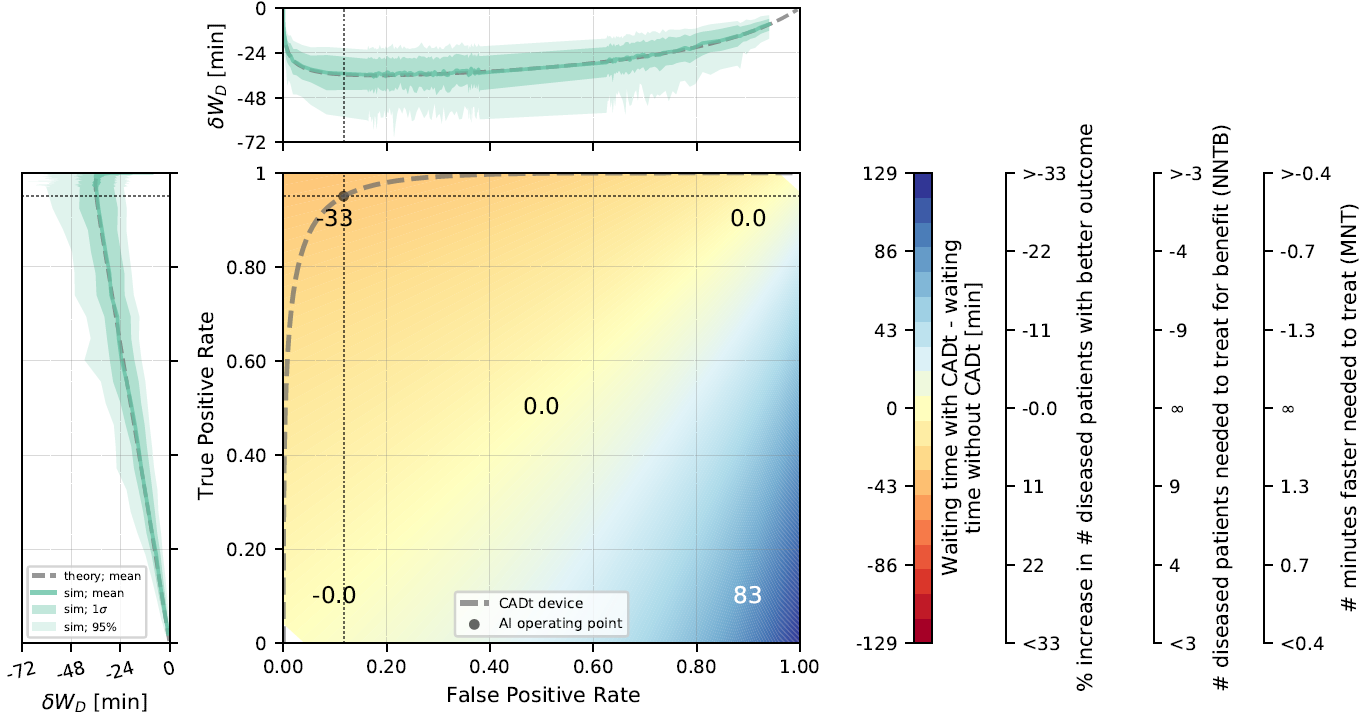}
\caption{A summary ROC plot for evaluating both the diagnostic and time-saving effectiveness of a CADt device.
The middle rainbow plot is an ROC space with an ROC curve (dashed dark gray) of a theoretical CADt device.
Color map represents theoretical mean time savings ($\delta W_D$) per diseased patient image, assuming a disease prevalence of 10\% in a relatively busy hospital (traffic intensity of 0.8) with only one radiologist and no emergent patient images.
The radiologist's average reading times for diseased and non-diseased patient images are both set at 10 minutes.
Positive $\delta W_D$ (blue region) means an overall time delay for diseased patient images, and negative $\delta W_D$ (red region) means an overall time savings.
The values printed on the color map are the $\delta W_D$'s at the corresponding points of false-positive and true-positive rates.
The dot represents the pre-determined AI operating point ($\text{Se} = 95\%, \text{Sp}=88\%$).
Top plot represents the theoretical $\delta W_D$ (dashed gray) along the ROC curve as a function of false-positive rate, and left plot represents the same theoretical $\delta W_D$ (dashed gray) along the curve but as a function of true-positive rate.
The green solid line represents the mean time savings along the ROC curve obtained from simulation.
The darker and lighter shaded areas indicate the 68\% and 95\% ranges from simulation around the mean time savings.
The black dotted vertical and horizontal lines indicate that the theoretical mean time-saving for diseased patients is roughly 36 minutes at the given operating point.
The color axis is translated to stroke patient outcome metrics based on Table 12 in Supplementary Content (Supplementary 2) of \cite{Saver2016}.}
\label{fig:SPIE_2D} 
\end{figure*}

For the purpose of evaluating a CADt device, we propose a summary plot as shown in Figure \ref{fig:SPIE_2D} based on Model B, describing both the diagnostic and time-saving effectiveness of a CADt device.
This plot is built upon a traditional receiver operating characteristic (ROC) analysis \citep{MetzNukeMed}, in which the ROC curve characterizes the diagnostic performance of the CADt device.
For a given radiologist workflow defined by a set of parameters, every point of False-Positive Rate (FPR) and True-Positive Rate (TPR) in the ROC space has an expected mean time savings per diseased patient image, $\delta W_D$, which is presented by the color map.
The device diagnostic performance is near ideal in the top left corner of the ROC space, where $\delta W_D$ is the most negative.

To show the time-saving effectiveness of a CADt device, $\delta W_D$ along the ROC curve is plotted as a function of FPR (top) and TPR (left).
At (FPR, TPR) = (0, 0), $\delta W_D$ is 0 minute because all images are classified as AI-negative i.e. no images are prioritized.
As both FPR and TPR increase along the ROC curve, the amount of time savings $|\delta W_D|$ increases since most AI-positive cases are truly diseased patient images.
As FPR and TPR continue to increase, the number of false-positive cases becomes dominant, reducing the device’s time-saving effectiveness.
When (FPR, TPR) = (1, 1), $\delta W_D$ goes back to 0 because all images are classified as AI-positive, and the system essentially has no priority classes. 

The mean time-savings for diseased patient images $\delta W_D$ can be directly linked to potential patient outcome.
For example, if our disease of interest is large vessel occlusion (LVO) stroke, $\delta W_D$ color axis on the right side of Figure \ref{fig:SPIE_2D} can be translated to three stroke patient outcome metrics.
According to Table 12 in Supplementary Content (Supplementary 2) of \cite{Saver2016}, for every 15 minutes sooner that a patient is treated, 3.9\% of stroke patients resulted in less disability.
This can be translated to the two other common LVO stroke patient outcome metrics - the number of patients needed to treat for benefit (NNTB) and the number of minutes faster needed to treat (MNT).
The relationships between $\delta W_D$ and LVO stroke patient outcome metrics are extrapolated linearly and shown in the three axes on the right side of Figure \ref{fig:SPIE_2D}.
As a result, the optimal $\delta W_D$ along the ROC curve is roughly -40 minutes, which corresponds to approximately 11\% increase in LVO stroke patients with less disability, more than 9 NNTB, and more than 1.4 MNT.  Remember that these results depend on our assumed reading rates and traffic intensity.  In the future we expect to gather clinical data to make more accurate estimates of reading rates, traffic intensity, and wait-time savings.

Based on our queueing approach, the time-saving effectiveness of a CADt device depends largely on the clinical settings.
Our model suggests that CADt devices with a typical AI diagnostic performance (95\% sensitivity and 89\% specificity) are most effective in a busy, short-staffed clinic.
All theoretical predictions agree with simulation results well within the 95\% confidence intervals.
All software used in making the theoretical calculations and the simulations in this paper will be made available on the Github site for the FDA's Division of Imaging, Diagnostics, and Software Reliability, https://github.com/DIDSR/QuCAD.

In this work where only one disease is considered, the CADt device is trained to identify the disease, and a patient image can either be diseased or non-diseased.
Under this consideration, when evaluating the time-saving effectiveness of the CADt, $\delta W_{\text{D}}$ is used as the performance metric because the CADt device is intended to benefit diseased patients with time critical conditions.
In the future, when we expand our work to a reading queue that consists of patient images with two or more diseases, a new performance matric will be defined to take into account other time-critical diseases that the CADt does not look for.

\section{Conclusion}
\label{sec:conclusion}

We present a mathematical framework based on queueing theory and the Recursive Dimensionality Reduction method to quantify the time-saving effectiveness of an AI-based medical device that prioritizes patient images based on its binary classification outputs.
Several models are developed to theoretically predict the wait-time-saving effectiveness of such a device as a function of various parameters, including disease prevalence, patient arrival rate, radiologist reading rate, number of radiologists on-site, AI sensitivity and specificity, as well as the presence of emergent patient images with the highest priority that overrides any AI prioritization.
The methodology proposed in this paper helps evaluate the time-saving performance of a CADt or any prioritization device.
The models presented here could also be used to evaluate discrimination algorithms in many other queueing contexts, such as serving customers or computer job queueing.
In the near future, we plan on expanding our model to clinical scenarios of multiple disease conditions, modalities, and anatomies with several CADt devices being used simultaneously.

\section{Acknowledgments}
\label{sec:acknowledgments}
The authors would like to thank Dr. Mor Harchol-Balter (harchol@cs.cmu.edu) and Dr. Takayuki Osogami (OSOGAMI@jp.ibm.com) for helping us understand their Recursive Dimensionality Reduction (RDR) method for complex queueing systems. In addition, the authors acknowledge funding from the Critical Path Program of the Center for Devices and Radiological Health. The authors also acknowledge funding by appointments to the Research Participation Program at the Center for Devices and Radiological Health administered by the Oak Ridge Institute for Science and Education through an interagency agreement between the U.S. Department of Energy and the U.S. Food and Drug Administration (FDA).

\bibliography{refs}

\appendix

\section{Markov Chain Matrices}
This appendix section provides the matrices involved for each of the four radiologist workflow models discussed in Section \ref{sec:models}. 

\subsection{Model A in with-CADt scenario}
\label{ECsec:queue_modelA_CADt}

Markov chain transition rate matrix $M_{\text{A}}$ is built upon Figure \ref{fig:simpleModel_transitionDiagram_lowHBOts}. 

\begin{equation}\label{eqn:simpleModel_transitionMatrix_HBOts}
M_{\text{A}} = 
 \left[
    \begin{array}{c|c|c|c|c}
      B_{00} & B_{01} & & & \\
      \hline
      B_{10} & A_{1} & A_{2} & &\\
      \hline
      & A_0 & A_{1} & A_{2} &\\
      \hline
      & & A_{0} & A_{1} & \ddots \\
      \hline
      & &      & \ddots & \ddots \\
    \end{array}
    \right],
\end{equation}
where 
\begin{align*}
\label{eqn:simpleModel_transitionMatrix_HBOts_As}
 A_0 &= \begin{pmatrix}
     \mu_- & 0 & 0 \\
     0 & 0 & 0 \\
     0 & 0 & 0
   \end{pmatrix}, \enspace
 A_1 = \begin{pmatrix}
      * & \lambda_+ & 0 \\
      t_1 & * & t_{12} \\
      t_2 & 0 & *
   \end{pmatrix}, \\ \\
 A_2 &=  \begin{pmatrix}
     \lambda_- & 0 & 0 \\
     0 & \lambda_- & 0 \\
     0 & 0 & \lambda_-
   \end{pmatrix}, \\ \\
 B_{01} &= \begin{pmatrix}   
      \lambda_- & 0 & 0 \\
      0 & \lambda_- & 0 \\
      0 & 0 & \lambda_-
   \end{pmatrix}, \enspace
 B_{00} = \begin{pmatrix}
      * & \lambda_+ & 0 \\
      t_1 & * & t_{12} \\
      t_2 & 0 & *
   \end{pmatrix}, \\ \\
 B_{10} &=  \begin{pmatrix}
     \mu_- & 0 & 0 \\
     0 & 0 & 0 \\
     0 & 0 & 0
   \end{pmatrix}. \numberthis
\end{align*}

$M_{\text{A}}$ had a tri-diagonal block structure defined by sub-matrices $A$s and $B$s, in which $*$'s are the negative of the sum of all elements in the corresponding row.
$B_{00}$, $B_{01}$, and $B_{10}$ are block matrices representing the boundary condition at the state of $n_- = 0$; states with $n_- < 0$ are forbidden because the reading queue cannot have a negative number for AI-negative patient images.
$A_{0}$, $A_{1}$, and $A_{2}$ are repetitive block structures that iterate along the diagonal axis of the matrix.

\subsection{Model B in with-CADt scenario}
\label{ECsec:queue_modelC_CADt}

\begin{figure*}[t]
\centering
\includegraphics[width=0.8\textwidth]{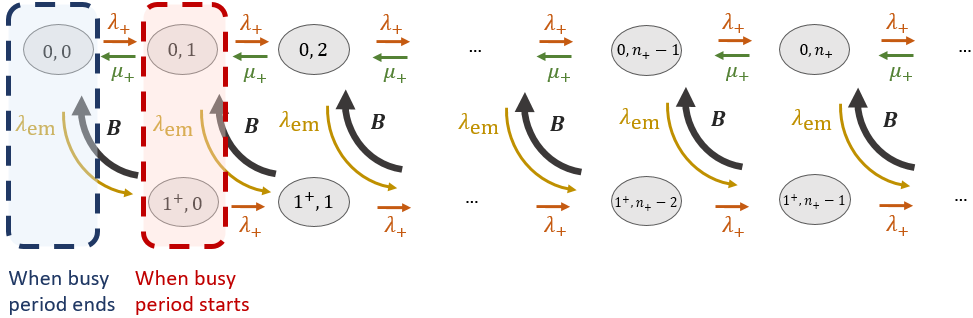}
\caption{The transition diagram to calculate inter-level passage times within the AI-positive (middle priority) in Model B in a with-CADt scenario.
The state is defined as $(n_{\text{em}}, n_+)$, and each column keeps track of $\ell = n_+ + \text{Min}(N_{\text{rad}}, n_{\text{em}}) + 1$.
Red and blue boxes represent the states at which the busy periods for the AI-negative patient images start and end respectively.}
\label{fig:modelC_transitionDiagram_lowHBO_interleveltime_Bs} 
\end{figure*}

This scenario has two busy periods ($B_1$ and $B_2$).
For each busy period, we first calculate its first three moments of the inter-level passage times using Figure \ref{fig:modelC_transitionDiagram_lowHBO_interleveltime_Bs}.
The states at which the two AI-negative busy periods start and end are highlighted.
For instance, \textbf{$B_1$} is the time period starting from $(0, 1)$ in red and ending at $(0, 0)$ in blue, regardless of any intermediate states that the system may go through.
The steps involved to calculate the first three moments are documented in Appendix A of \cite{HarcholBalter2005}.

Based on Figure \ref{fig:modelC_transitionDiagram_lowHBO_interleveltime_Bs}, the transition probability matrix $P_{\text{B}}$ is given below.
\begin{equation}\label{eqn:modelC_transitionProbMatrix_lowHBO_interleveltime}
P_{\text{B}} = 
 \left[
    \begin{array}{c|c|c|c|c}
      \mathcal{L}_{1} & \mathcal{F}_{1} & & &\\
      \hline
      \mathcal{B}_{2} & \mathcal{L}_{2} & \mathcal{F}_{2} & &\\
      \hline
      & \mathcal{B}_{3} & \mathcal{L}_{3} & \mathcal{F}_{3} & \\
      \hline
      & & \mathcal{B}_{4} & \mathcal{L}_{4} & \ddots\\
      \hline
      & &      & \ddots & \ddots \\
    \end{array}
    \right],
\end{equation}
where 
\begin{align*}
 \mathcal{B}_{\ell=2} &= \begin{pmatrix}
     \frac{\mu_+}{\lambda_{\text{em}}+\lambda_++\mu_+}\\[6pt]
     \frac{t_1}{\lambda_++t_1+t_{12}} \\[6pt]
     \frac{t_2}{\lambda_++t_{2}}\\
   \end{pmatrix}, \\ \\
 \mathcal{B}_{\ell\geq 3} &= \begin{pmatrix}
     \frac{\mu_+}{\lambda_{\text{em}}+\lambda_++\mu_+} & 0 & 0\\[6pt]
     \frac{t_1}{\lambda_++t_1+t_{12}} & 0 & 0 \\[6pt]
     \frac{t_2}{\lambda_++t_{2}} & 0 & 0 \\
   \end{pmatrix}, \\ \\
 \mathcal{L}_{\ell=1} &= \begin{pmatrix}
      0
   \end{pmatrix}, \enspace
 \mathcal{L}_{\ell\geq 2} = \begin{pmatrix}
      0 & 0 & 0 \\
      0 & 0 & \frac{t_{12}}{\lambda_++t_1+t_{12}} \\
      0 & 0 & 0 \\
   \end{pmatrix}, 
 \\ \\
 \mathcal{F}_{\ell=1} &= \begin{pmatrix}
     \frac{\lambda_+}{\lambda_+ + \lambda_{\text{em}}} & \frac{\lambda_{\text{em}}}{\lambda_+ + \lambda_{\text{em}}} & 0
   \end{pmatrix}, 
 \\ \\
 \mathcal{F}_{\ell\geq 2} &= \begin{pmatrix}
     \frac{\lambda_+}{\lambda_+ + \lambda_{\text{em}}+\mu_+} & \frac{\lambda_{\text{em}}}{\lambda_+ + \lambda_{\text{em}}+\mu_+} & 0 \\
     0 & \frac{\lambda_+}{\lambda_+ +t_1+t_{12}} & 0 \\
     0 & 0 & \frac{\lambda_+}{\lambda_+ +t_2} \\
   \end{pmatrix}. \numberthis
\end{align*}
Here, the $t$-parameters are the approximated exponential rates from the transition $\textbf{B}$ in Figure \ref{fig:modelC_transitionDiagram_lowHBO_interleveltime_Bs}.
With $\mathcal{F}$, $\mathcal{L}$, and $\mathcal{B}$, \cite{HarcholBalter2005} provide the framework to obtain the \textbf{$G$} matrix, which contains the probabilities of the busy periods involved, and the \textbf{$Z_r$} matrices, which have the $r$-th moments of the busy periods.
For the AI-negative priority class in Model B, the \textbf{$Z_r$} matrix has a dimension of $3\times1$, where the first and second elements are the $r$-th moments of \textbf{$B_1$} and \textbf{$B_2$} respectively.
For each of the two busy periods, a two-phase Coxian distribution can be used to approximate the distribution shape using Equation \ref{eqn:simpleModel_HBOts}.

Let $t^{(1)}_1$, $t^{(1)}_{12}$, and $t^{(1)}_2$ be the approximated rates for \textbf{$B_1$}, and $t^{(2)}_1$, $t^{(2)}_{12}$, and $t^{(2)}_2$ be the approximated rates for \textbf{$B_2$}.
The transition rate matrix $M_{\text{B}}$ for Figure \ref{fig:modelC_transitionDiagram_lowHBO} is given below.
\begin{equation}\label{eqn:modelC_transitionMatrix_HBOts}
M_{\text{B}} = 
 \left[
    \begin{array}{c|c|c|c}
      B_{00} & B_{01} & &\\
      \hline
      B_{10} & A_{1} & A_{2} & \\
      \hline
      & A_0 & A_{1} & \ddots\\
      \hline
      & & A_{0} & \ddots \\
      \hline
      & &  & \ddots \\
    \end{array}
    \right],
\end{equation}
where 
\begin{align*}
 A_0 &= \begin{pmatrix}
     \mu_- & 0 & 0 & 0 & 0\\
     0 & 0 & 0 & 0 & 0\\
     0 & 0 & 0 & 0 & 0\\
     0 & 0 & 0 & 0 & 0
   \end{pmatrix}, 
 \\ \\
 A_1 &= \begin{pmatrix}
      * & \lambda_+ & 0 & \lambda_{\text{em}} & 0\\
      t^{(1)}_1 & * & t^{(1)}_{12} & 0 & 0\\
      t^{(1)}_2 & 0 & * & 0 & 0 \\
      t^{(2)}_1 & 0 & 0 & * & t^{(2)}_{12}\\
      t^{(2)}_2 & 0 & 0 & 0 & * \\
   \end{pmatrix},
\\ \\
 A_2 &= \begin{pmatrix}
     \lambda_- & 0 & 0 & 0 & 0 \\
     0 & \lambda_- & 0 & 0 & 0 \\
     0 & 0 & \lambda_- & 0 & 0 \\
     0 & 0 & 0 & \lambda_- & 0 \\
     0 & 0 & 0 & 0 & \lambda_- \\
   \end{pmatrix}, \\ \\
 B_{00} &= \begin{pmatrix}
      * & \lambda_+ & 0 & \lambda_{\text{em}} & 0\\
      t^{(1)}_1 & * & t^{(1)}_{12} & 0 & 0\\
      t^{(1)}_2 & 0 & * & 0 & 0 \\
      t^{(2)}_1 & 0 & 0 & * & t^{(2)}_{12}\\
      t^{(2)}_2 & 0 & 0 & 0 & * \\
   \end{pmatrix}, \\ \\
 B_{01} &= \begin{pmatrix}
     \lambda_- & 0 & 0 & 0 & 0 \\
     0 & \lambda_- & 0 & 0 & 0 \\
     0 & 0 & \lambda_- & 0 & 0 \\
     0 & 0 & 0 & \lambda_- & 0 \\
     0 & 0 & 0 & 0 & \lambda_- \\
   \end{pmatrix}, \\ \\
 B_{10} &= \begin{pmatrix}
     \mu_- & 0 & 0 & 0 & 0\\
     0 & 0 & 0 & 0 & 0\\
     0 & 0 & 0 & 0 & 0\\
     0 & 0 & 0 & 0 & 0
   \end{pmatrix}.\numberthis \\
\end{align*}

The sub-matrices in $M_{\text{B}}$ are very similar to that in $M_{\text{A}}$ (Equations \ref{eqn:simpleModel_transitionMatrix_HBOts_As} and \ref{eqn:simpleModel_transitionMatrix_HBOts_Bs}).
The one difference is that these sub-matrices are now $5\times5$ instead of $3\times3$ due to the extra row of truncated states in Figure \ref{fig:modelC_transitionDiagram_lowHBO} compared to Figure \ref{fig:simpleModel_transitionDiagram_lowHBO}.

\subsection{Model C in without-CADt scenario}
\label{ECsec:queue_modelD_noCADt}

The transition rate matrix $M_{\text{C}_{\text{noCADt}}}$ is built upon Figure \ref{fig:modelD_transitionDiagram_noCADt}.
\begin{equation}\label{eqn:modelD_transitionMatrix_noCADt}
M_{\text{C}_{\text{noCADt}}} = 
 \left[
    \begin{array}{c|c|c|c|c}
      B_{00} & B_{01} & & & \\
      \hline
      B_{10} & A_{1} & A_{2} & &\\
      \hline
      & A_0 & A_{1} & A_{2} & \\
      \hline
      & & A_{0} & A_{1} & \ddots\\
      \hline
      & &  & \ddots & \ddots\\
    \end{array}
    \right],
\end{equation}
where 
\begin{align*}
 A_0 &= \begin{pmatrix}
     2\mu_{\text{nonEm}} & 0 & 0 & 0\\
     0 & \mu_{\text{nonEm}} & 0 & 0\\
     0 & 0 & 0 & 0\\
     0 & 0 & 0 & 0
   \end{pmatrix}, \\ \\
 A_1 &= \begin{pmatrix}
      * & \lambda_{\text{em}} & 0 & 0\\
      \mu_{\text{em}} & * & \lambda_{\text{em}} & 0\\
      0 & t_1 & * & t_{12} \\
      0 & t_2 & 0 & *
   \end{pmatrix}, \\ \\
 A_2 &= \begin{pmatrix}
     \lambda_{\text{nonEm}} & 0 & 0 & 0 \\
     0 & \lambda_{\text{nonEm}} & 0 & 0 \\
     0 & 0 & \lambda_{\text{nonEm}} & 0 \\
     0 & 0 & 0 & \lambda_{\text{nonEm}} 
   \end{pmatrix}, 
\end{align*}

\begin{align*}
 B_{00} &= \scalemath{0.7}{
    \begin{pmatrix}[cccc:cccc]
      * & \lambda_{\text{em}} & 0 & 0 & \lambda_{\text{nonEm}} & 0 & 0 & 0 \\
      \mu_+ & * & \lambda_{\text{em}} & 0 & 0 & \lambda_{\text{nonEm}} & 0 & 0 \\
      0 & t_1 & * & t_{12} & 0 & 0 & \lambda_{\text{nonEm}} & 0 \\
      0 & t_2 & 0 & * & 0 & 0 & 0 & \lambda_{\text{nonEm}} \\
      \hdashline[2pt/2pt]
      \mu_{\text{nonEm}} & 0 & 0 & 0 & * & \lambda_{\text{em}} & 0 & 0 \\
      0 & \mu_{\text{nonEm}} & 0 & 0 & \mu & * & \lambda_{\text{em}} & 0\\
      0 & 0 & 0 & 0 & 0 & t_1 & * & t_{12} \\
      0 & 0 & 0 & 0 & 0 & t_2 & 0 & * \\
   \end{pmatrix}}, \\ \\
 B_{01} &=  \scalemath{0.8} {\begin{pmatrix}   
      0 & 0& 0& 0 \\
      0 & 0& 0& 0 \\
      0 & 0& 0& 0 \\
      0 & 0& 0& 0 \\
      \hdashline[2pt/2pt]
      \lambda_{\text{nonEm}} & 0 & 0 & 0\\
      0 & \lambda_{\text{nonEm}} & 0 & 0\\
      0 & 0 & \lambda_{\text{nonEm}} & 0 \\
      0 & 0 & 0 & \lambda_{\text{nonEm}}\\
   \end{pmatrix}}, \\ \\
 B_{10} &=   \scalemath{1}{\begin{pmatrix}[cccc:cccc]
     0 & 0& 0& 0 & 2\mu_{\text{nonEm}} & 0 & 0 & 0\\
     0 & 0& 0& 0 & 0 & \mu_{\text{nonEm}} & 0 & 0\\
     0 & 0& 0& 0 & 0 & 0 & 0 & 0\\
     0 & 0& 0& 0 & 0 & 0 & 0 & 0
   \end{pmatrix}}. \numberthis \\ 
\end{align*}

\subsection{Model C in with-CADt scenario}
\label{ECsec:queue_modelD_CADt}

\begin{figure*}[t]
\centering
\includegraphics[width=0.85\textwidth]{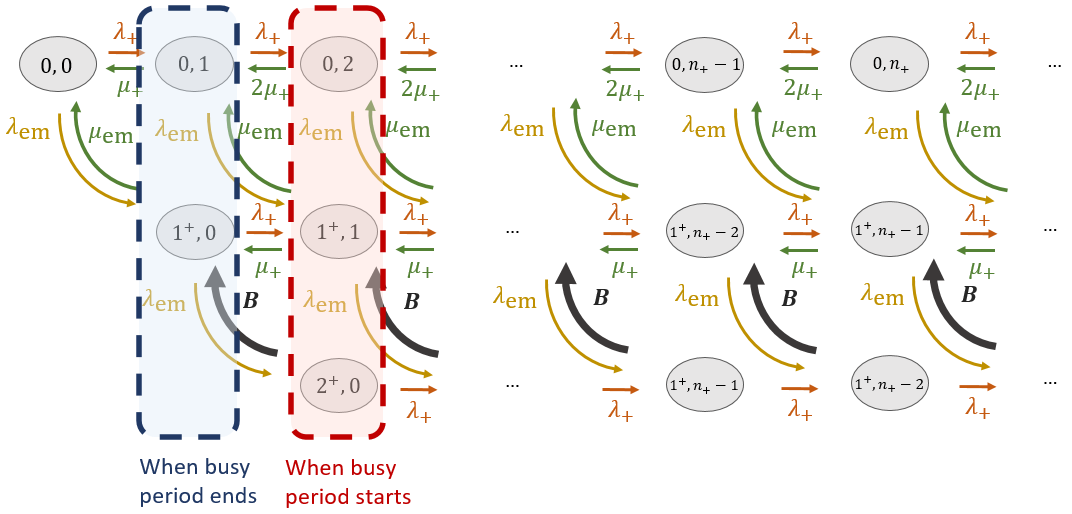}
\caption{The transition diagram to calculate inter-level passage times within the AI-positive (middle-priority) class in Model C in a with-CADt scenario.}
\label{fig:modelD_transitionDiagram_lowHBO_interleveltime_Bs} 
\end{figure*}

This scenario has six busy periods ($B_1$ to $B_6$).
For each busy period, we first calculate its conditional probability and the first three moments of the inter-level passage times using Figure \ref{fig:modelD_transitionDiagram_lowHBO_interleveltime_Bs}.
And the corresponding transition probability matrix $P_{\text{C}}$ is given below.

\begin{equation}\label{eqn:modelD_transitionProbMatrix_lowHBO_interleveltime}
P_{\text{C}} = 
 \left[
    \begin{array}{c|c|c|c|c}
      \mathcal{L}_{1} & \mathcal{F}_{1} & & &\\
      \hline
      \mathcal{B}_{2} & \mathcal{L}_{2} & \mathcal{F}_{2} & &\\
      \hline
      & \mathcal{B}_{3} & \mathcal{L}_{3} & \mathcal{F}_{3} &  \\
      \hline
      & & \mathcal{B}_{4} & \mathcal{L}_{4} & \ddots\\
      \hline
      & &  & \ddots & \ddots\\
    \end{array}
    \right],
\end{equation}
where 
\allowdisplaybreaks
\begin{align*}
    \mathcal{B}_{\ell=2} &= \scalemath{1} {\begin{pmatrix}
     \frac{\mu_+}{\lambda_{\text{em}}+\lambda_++\mu_+}\\[6pt]
     \frac{\mu_{\text{em}}}{\lambda_{\text{em}}+\lambda_++\mu_{\text{em}}} 
        \end{pmatrix}} , \\ \\
    \mathcal{B}_{\ell=3} &= \scalemath{1} {\begin{pmatrix}
     \frac{2\mu_+}{\lambda_{\text{em}}+\lambda_++2\mu_+} & 0\\[6pt]
     \frac{\mu_{\text{em}}}{\lambda_{\text{em}}+\lambda_++\mu_++\mu_{\text{em}}} & \frac{\mu_+}{\lambda_{\text{em}}+\lambda_++\mu_++\mu_{\text{em}}} \\[6pt]
     0 & \frac{t_1}{\lambda_++t_1+t_{12}} \\[6pt]
     0& \frac{t_2}{\lambda_++t_{2}}\\
    \end{pmatrix}} \\ \\
    \mathcal{B}_{\ell\geq4} &= \scalemath{1}{\begin{pmatrix}
     \frac{2\mu_+}{\lambda_{\text{em}}+\lambda_++2\mu_+} & 0 & 0 & 0\\[6pt]
     \frac{\mu_{\text{em}}}{\lambda_{\text{em}}+\lambda_++\mu_++\mu_{\text{em}}} &  \frac{\mu_+}{\lambda_{\text{em}}+\lambda_++\mu_++\mu_{\text{em}}} & 0 & 0\\[6pt]
     0 & \frac{t_1}{\lambda_++t_1+t_{12}} & 0 & 0\\[6pt]
     0& \frac{t_2}{\lambda_++t_{2}} & 0 & 0\\
    \end{pmatrix}},
    \\ \\
    \mathcal{L}_{\ell=1} &= \begin{pmatrix}
      0
    \end{pmatrix}, \enspace
    \mathcal{L}_{\ell=2} = \begin{pmatrix}
      0 & 0 \\
      0 & 0
    \end{pmatrix}, \\ \\
    \mathcal{L}_{\ell\geq3} &= \begin{pmatrix}
      0 & 0 & 0 & 0 \\
      0 & 0 & 0 & 0 \\
      0 & 0 & 0 & \frac{t_{12}}{\lambda_++t_1+t_{12}} \\
      0 & 0 & 0 & 0
    \end{pmatrix},
    \\ \\
    \mathcal{F}_{\ell=1} &= \scalemath{1}{\begin{pmatrix}
     \frac{\lambda_+}{\lambda_{\text{em}}+\lambda_+} & \frac{\lambda_{\text{em}}}{\lambda_{\text{em}}+\lambda_+}
    \end{pmatrix}}, \\ \\
    \mathcal{F}_{\ell=2} &= \scalemath{1}{\begin{pmatrix}
     \frac{\lambda_+}{\lambda_{\text{em}}+\lambda_++\mu_+} & \frac{\lambda_{\text{em}}}{\lambda_{\text{em}}+\lambda_++\mu_+} & 0 & 0 \\
     0 & \frac{\lambda_+}{\lambda_{\text{em}}+\lambda_++\mu_{\text{em}}} & \frac{\lambda_{\text{em}}}{\lambda_{\text{em}}+\lambda_++\mu_{\text{em}}} & 0
    \end{pmatrix}},
    \\ \\
    \mathcal{F}_{\ell\geq3} & = \scalemath{0.8}{\begin{pmatrix}
     \frac{\lambda_+}{\lambda_{\text{em}}+\lambda_++2\mu_+} & \frac{\lambda_{\text{em}}}{\lambda_{\text{em}}+\lambda_++2\mu_+} & 0 & 0 \\
     0 & \frac{\lambda_+}{\lambda_{\text{em}}+\lambda_++\mu_++\mu_{\text{em}}} & \frac{\lambda_{\text{em}}}{\lambda_{\text{em}}+\lambda_++\mu_++\mu_{\text{em}}} & 0 \\
     0 & 0 & \frac{\lambda_+}{\lambda_++t_1+t_{12}} & 0\\
     0 & 0 & 0 & \frac{\lambda_+}{\lambda_++t_2}
    \end{pmatrix}}. \\ \numberthis
\end{align*}

With $P_{\text{C}}$, the conditional probabilities and first three moments of inter-level passage times for all six busy periods are computed according to Appendix A of \cite{HarcholBalter2005}.
For most busy periods, three $t$-parameters are sufficient for the approximation.
However, for $B_2$ and $B_5$, due to the two extra Erlang phases, two additional parameters $t_0$ and $t_{01}$ are required.

Let $t^{(i)}_j$ denote the $t_j$-parameter for a busy period $B_i$.
The transition rate matrix $M_{C_{\text{CADt}}}$ for the AI-negative, lowest-priority class from Figure \ref{fig:modelD_transitionDiagram_lowHBO} is given by
\begin{equation}\label{eqn:modelD_transitionMatrix_CADt}
M_{\text{C}_{\text{CADt}}} = 
 \left[
    \begin{array}{c|c|c|c}
      B_{00} & B_{01} & & \\
      \hline
      B_{10} & A_{1} & A_{2} & \\
      \hline
      & A_0 & A_{1} & \ddots \\
      \hline
      & & A_{0} & \ddots \\
      \hline
      & &  & \ddots \\
    \end{array}
    \right].
\end{equation}
All $A$ sub-matrices are $17\times17$. 
Here, $\mathbb{0}_{14}$ denotes a $14\times14$ zero matrix, and $\mathbb{I}_{17}$ is a $17\times17$ identity matrix.
\begin{align*} 
 A_0 &= \begin{pmatrix}
     2\mu_- &       &       &\\
            & \mu_- &       &\\
            &       & \mu_- & \\
            &       &       & \mathbb{0}_{14}
   \end{pmatrix}, \\ \\
 A_2 &= \lambda_- \mathbb{I}_{17}, \\ \\
 A_1&= \scalemath{0.75}{\begin{pmatrix}[ccc:c:c:c:c:c:c]
        * & \lambda_+ & \lambda_{\text{em}} & & & & & & \\
        \mu_+ & * & & \mathbf{p_1}\lambda_{+} & \mathbf{p_2}\lambda_{+} & \mathbf{p_3}\lambda_{\text{em}} & \mathbf{p_4}\lambda_{\text{em}} & & \\
        \mu_{\text{em}} & & * & & & \mathbf{p_3}\lambda_+ & \mathbf{p_4}\lambda_+ & \mathbf{p_5}\lambda_{\text{em}} & \mathbf{p_6}\lambda_{\text{em}}\\[6pt]
        \hdashline[2pt/2pt]
        & \mathbf{t}^{(1)} & & \mathbb{T}^{(1)}_4 & & & & & \\[6pt]
        \hdashline[2pt/2pt]
        & & \mathbf{t}^{(2)} & & \mathbb{T}^{(2)}_5 & & & & \\[6pt]
        \hdashline[2pt/2pt]
        & \mathbf{t}^{(3)} & & & & \mathbb{T}^{(3)}_6 & & &\\[6pt]
        \hdashline[2pt/2pt]
        & & \mathbf{t}^{(4)} & & & & \mathbb{T}^{(4)}_7 & & \\[6pt]
        \hdashline[2pt/2pt]
        & \mathbf{t}^{(5)} & & & & & & \mathbb{T}^{(5)}_8 &\\[6pt]
        \hdashline[2pt/2pt]
        & & \mathbf{t}^{(6)} & & & & & & \mathbb{T}^{(6)}_9
   \end{pmatrix}},  \\ \\ \numberthis 
\end{align*} 
where, for $i = 1, 3, 4, 6$,
\begin{align*}
 \mathbf{p_i} = \begin{pmatrix}
    p_i & 0
 \end{pmatrix}, \enspace
 \mathbf{t}^{(i)} = \begin{pmatrix}
    t_1^{(i)} \\[6pt] t_2^{(i)}  
 \end{pmatrix},  \enspace
  \mathbb{T}^{(i)}_k = \begin{pmatrix}
    * & t_{12}^{(i)} \\[6pt]
     & *
 \end{pmatrix}.
\end{align*}
For $i = 2, 5$, because of the extra Erlang phase, the sub-matrices have an extra row and/or column.
\begin{align*}
 \mathbf{p_i} = \begin{pmatrix}
    p_i & 0 & 0
 \end{pmatrix}, \enspace
 \mathbf{t}^{(i)} = \begin{pmatrix}
    t_0^{(i)} \\[6pt] t_1^{(i)} \\[6pt] t_2^{(i)}  
 \end{pmatrix},  \enspace
  \mathbb{T}^{(i)}_k = \begin{pmatrix}
   * & t_{01}^{(i)} &  \\[6pt]
    & * & t_{12}^{(i)} \\[6pt]
    &  & *
 \end{pmatrix}.
\end{align*}
Similarly, the boundary $B$ sub-matrices are given below.
\begin{align*}
 B_{01} &= \begin{pmatrix}
     \mathbb{0}_{17} \\[6pt]
     \hdashline[2pt/2pt]
     \lambda_- \mathbb{I}_{17}
   \end{pmatrix}, \enspace
 B_{10} = \begin{pmatrix}[c:c]
    \mathbb{0}_{17} & \begin{matrix} 2\mu_+ & & & \\ & \mu_+ & & \\ & & \mu_+ & \\ & & & \mathbb{0}_{14} \end{matrix}
   \end{pmatrix}, \\ \\
 B_{00} &= \begin{pmatrix}[c:c]
    A_1 & \lambda_- \mathbb{I}_{17} \\[6pt]
     \hdashline[2pt/2pt]
     \begin{matrix} \mu_+ & & & \\ & \mu_+ & & \\ & & \mu_+ & \\ & & & \mathbb{0}_{14} \end{matrix} & A_1
   \end{pmatrix}. \numberthis 
\end{align*}

\subsection{Model D in without-CADt scenario}
\label{ECsec:queue_modelE_noCADt}

The transition rate matrix $M_{\text{D}_{\text{noCADt}}}$ for Figure \ref{fig:modelE_transitionDiagram_noCADt} is given below. 
\begin{equation}\label{eqn:modelE_transitionMatrix_noCADt}
M_{\text{D}_{\text{noCADt}}} = 
 \left[
    \begin{array}{c|c|c|c|c}
      B_{00} & B_{01} & & & \\
      \hline
      B_{10} & A_{1} & A_{2} & &\\
      \hline
      & A_0 & A_{1} & A_{2} &\\
      \hline
      & & A_{0} & A_{1} & \ddots \\
      \hline
      & &  & \ddots & \ddots \\
    \end{array}
    \right],
\end{equation}
where 
\begin{align*}
 A_0 &= \begin{pmatrix}
     \pi\mu_{D} & (1-\pi)\mu_{D} & 0 & 0 & 0 & 0\\
     \pi\mu_{ND} & (1-\pi)\mu_{ND} & 0 & 0 & 0 & 0 \\
     0 & 0 & 0 & 0 & 0 & 0\\
     0 & 0 & 0 & 0 & 0 & 0\\
     0 & 0 & 0 & 0 & 0 & 0\\
     0 & 0 & 0 & 0 & 0 & 0
   \end{pmatrix}, \\ \\
 A_1 &= \begin{pmatrix}
      * & 0 & \lambda_{\text{em}} & 0 & 0 & 0\\
      0 & * & 0 & 0 & \lambda_{\text{em}} & 0\\
      t_1 & 0 & * & t_{12} & 0 & 0 \\
      t_2 & 0 & 0 & * & 0 & 0 \\
      0 & t_1 & 0 & 0 & * & t_{12}\\
      0 & t_2 & 0 & 0 & 0 & * \\
   \end{pmatrix}, \\ \\
 A_2 &= \lambda_{\text{nonEm}} \mathbb{I}_{6}, \\ \\
 B_{00} &= \begin{pmatrix}
      * & \lambda_{\text{em}} & 0 \\
      t_1 & * & t_{12} \\
      t_2 & 0 & *
   \end{pmatrix}, \enspace
 B_{10} = \begin{pmatrix}
     \mu_{D} & 0 & 0 \\
     \mu_{ND} & 0 & 0 \\
     0 & 0 & 0 \\
     0 & 0 & 0 \\
     0 & 0 & 0 \\
     0 & 0 & 0
   \end{pmatrix}, \\ \\
 B_{01} &= \scalemath{0.75}{\begin{pmatrix}   
      \pi\lambda_{\text{nE}} & (1-\pi)\lambda_{\text{nE}} & 0 & 0 & 0 & 0 \\
      0 & 0 & \pi\lambda_{\text{nE}} & 0 & (1-\pi)\lambda_{\text{nE}} & 0 \\
      0 & 0 & 0 & \pi\lambda_{\text{nE}} & 0 & (1-\pi)\lambda_{\text{nE}}
   \end{pmatrix}}, \\ \numberthis
\end{align*}
where $\lambda_{\text{nE}}$ refers to the arrival rate of non-emergent subgroup. Note that both $M_{\text{D}_{\text{noCADt}}}$ and $M_\text{A}$ (Equation \ref{eqn:simpleModel_transitionMatrix_HBOts}) describe a queueing system with two priority classes and one radiologist.
However, because $\mu_D \neq \mu_{ND}$, the size of $A$ sub-matrices grow from $3\times3$ to $6\times6$ due to the extra $i$ in the state definition and the extra set of truncated states to keep track of disease status of the interrupted case.

\subsection{Model D in with-CADt scenario}
\label{ECsec:queue_modelE_CADt}

\begin{figure*}[t]
\centering
\includegraphics[width=0.85\textwidth]{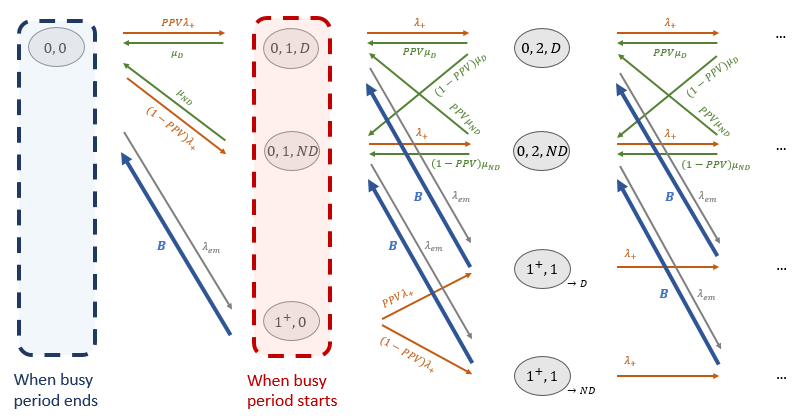}
\caption{The transition diagram to calculate inter-level passage time within the AI-positive (middle-priority) class in Model D in a with-CADt scenario.}
\label{fig:modelE_transitionDiagram_lowHBO_interleveltime_Bs}  
\end{figure*}

This scenario has three busy periods ($B_1$ to $B_3$).
For each busy period, we calculate its conditional probability and the first three moments of the inter-level passage times using Figure \ref{fig:modelE_transitionDiagram_lowHBO_interleveltime_Bs} and the corresponding transition probability matrix $P_{\text{D}}$.

\begin{equation}\label{eqn:modelE_transitionProbMatrix_lowHBO_interleveltime}
P_{\text{E}} = 
 \left[
    \begin{array}{c|c|c|c|c}
      \mathcal{L}_{1} & \mathcal{F}_{1} & & &\\
      \hline
      \mathcal{B}_{2} & \mathcal{L}_{2} & \mathcal{F}_{2} & &\\
      \hline
      & \mathcal{B}_{3} & \mathcal{L}_{3} & \mathcal{F}_{3} &  \\
      \hline
      & & \mathcal{B}_{4} & \mathcal{L}_{4} & \ddots\\
      \hline
      & &              & \ddots & \ddots\\
    \end{array}
    \right],
\end{equation}
where 
\begin{align*}
    \mathcal{B}_{\ell=2} &= \scalemath{1} {\begin{pmatrix}
     \frac{\mu_D}{\lambda_{\text{em}}+\lambda_++\mu_D}\\[6pt]
     \frac{\mu_{ND}}{\lambda_{\text{em}}+\lambda_++\mu_{ND}}\\[6pt]
     \frac{t_1}{\lambda_++t_1+t_{12}}\\[6pt]
     \frac{t_2}{\lambda_++t_2} 
        \end{pmatrix}} , \\ \\
    \mathcal{B}_{\ell=3} &= \scalemath{1} {\begin{pmatrix}
     \frac{PPV\mu_D}{\lambda_{\text{em}}+\lambda_++\mu_D} &
     \frac{(1-PPV)\mu_D}{\lambda_{\text{em}}+\lambda_++\mu_D} & 0 & 0 \\[6pt]
     \frac{PPV\mu_{ND}}{\lambda_{\text{em}}+\lambda_++\mu_{ND}} & \frac{(1-PPV)\mu_{ND}}{\lambda_{\text{em}}+\lambda_++\mu_{ND}} & 0 & 0\\[6pt]
     \frac{t_1}{\lambda_++t_1+t_{12}} & 0 & 0 & 0 \\[6pt]
     \frac{t_2}{\lambda_++t_2} & 0 & 0 & 0 \\[6pt]
     0 & \frac{t_1}{\lambda_++t_1+t_{12}} & 0 & 0 \\[6pt]
     0 & \frac{t_2}{\lambda_++t_{2}} & 0 & 0 
    \end{pmatrix}} \\ \\
    \mathcal{B}_{\ell\geq4} &= \scalemath{1}{\begin{pmatrix}
     \frac{PPV\mu_D}{\lambda_{\text{em}}+\lambda_++\mu_D} &
     \frac{(1-PPV)\mu_D}{\lambda_{\text{em}}+\lambda_++\mu_D} & 0 & 0 & 0 & 0 \\[6pt]
     \frac{PPV\mu_{ND}}{\lambda_{\text{em}}+\lambda_++\mu_{ND}} & 
     \frac{(1-PPV)\mu_{ND}}{\lambda_{\text{em}}+\lambda_++\mu_{ND}} & 0 & 0 & 0 & 0 \\[6pt]
     \frac{t_1}{\lambda_++t_1+t_{12}} & 0 & 0 & 0 & 0 & 0 \\[6pt]
     \frac{t_2}{\lambda_++t_{2}} & 0 & 0 & 0 & 0 & 0 \\[6pt]
     0 & \frac{t_1}{\lambda_++t_1+t_{12}} & 0 & 0 & 0 & 0\\[6pt]
     0 & \frac{t_2}{\lambda_++t_{2}} & 0 & 0 & 0 & 0\\
    \end{pmatrix}},
    \\ \\
    \mathcal{L}_{\ell=1} &= \scalemath{1} {\begin{pmatrix}
      0
    \end{pmatrix}}, \enspace
    \mathcal{L}_{\ell=2} = \scalemath{1}{\begin{pmatrix}
      0 & 0 & 0 & 0\\
      0 & 0 & 0 & 0\\
      0 & 0 & 0 & 0\\
      0 & 0 & 0 & 0
    \end{pmatrix}}, \\ \\
    \mathcal{L}_{\ell\geq3} &= \scalemath{1}{\begin{pmatrix}
      0 & 0 & 0 & 0 & 0 & 0 \\
      0 & 0 & 0 & 0 & 0 & 0 \\
      0 & 0 & 0 & \frac{t_{12}}{\lambda_++t_1+t_{12}} & 0 & 0  \\
      0 & 0 & 0 & 0 & 0 & 0 \\
      0 & 0 & 0 & 0 & 0 & \frac{t_{12}}{\lambda_++t_1+t_{12}} \\
      0 & 0 & 0 & 0 & 0 & 0 \\
    \end{pmatrix}},
    \\ \\
    \mathcal{F}_{\ell=1} &= \scalemath{1}{\begin{pmatrix}
     \frac{PPV\lambda_+}{\lambda_{\text{em}}+\lambda_+} & \frac{(1-PPV)\lambda_+}{\lambda_{\text{em}}+\lambda_+} & \frac{\lambda_{\text{em}}}{\lambda_{\text{em}}+\lambda_+} & 0
    \end{pmatrix}}, 
\end{align*}

\begin{strip}
\centering
\begin{align*}
    \mathcal{F}_{\ell=2} &= \scalemath{1}{\begin{pmatrix}
     \frac{\lambda_+}{\lambda_{\text{em}}+\lambda_++\mu_D} & 0 & \frac{\lambda_{\text{em}}}{\lambda_{\text{em}}+\lambda_++\mu_D} & 0 & 0 & 0 \\
     0 & \frac{\lambda_+}{\lambda_{\text{em}}+\lambda_++\mu_{ND}} & 0 & 0 & \frac{\lambda_{\text{em}}}{\lambda_{\text{em}}+\lambda_++\mu_{ND}} & 0 \\
     0 & 0 & \frac{PPV\lambda_+}{\lambda_++t_1+t_{12}} & 0 & 
     \frac{(1-PPV)\lambda_+}{\lambda_++t_1+t_{12}} & 0 \\
     0 & 0 & 0 & \frac{PPV\lambda_+}{\lambda_++t_2} & 0 & 
     \frac{(1-PPV)\lambda_+}{\lambda_++t_2} 
    \end{pmatrix}},
    \\ \\
    \mathcal{F}_{\ell\geq3} & = \scalemath{1}{\begin{pmatrix}
     \frac{\lambda_+}{\lambda_{\text{em}}+\lambda_++\mu_D} & 0 & \frac{\lambda_{\text{em}}}{\lambda_{\text{em}}+\lambda_++\mu_D} & 0 & 0 & 0 \\
     0 & \frac{\lambda_+}{\lambda_{\text{em}}+\lambda_++\mu_{ND}} & 0 & 0 & \frac{\lambda_{\text{em}}}{\lambda_{\text{em}}+\lambda_++\mu_{ND}} & 0 \\
     0 & 0 & \frac{\lambda_+}{\lambda_++t_1+t_{12}} & 0 & 0 & 0\\
     0 & 0 & 0 & \frac{\lambda_+}{\lambda_++t_2} & 0 & 0 \\
     0 & 0 & 0 & 0 & \frac{\lambda_+}{\lambda_++t_1+t_{12}} & 0 \\
     0 & 0 & 0 & 0 & 0 & \frac{\lambda_+}{\lambda_++t_2}
    \end{pmatrix}}. \numberthis
\end{align*}
\end{strip}
Each of the three busy periods has a set of $t$-parameters (Equation \ref{eqn:simpleModel_HBOts}) approximated from a two-phase Coxian distribution.

Let $t^{(i)}_j$ denote the $t_j$ parameter for the busy period $B_i$.
The transition rate matrix $M_{\text{D}_{\text{CADt}}}$ for the AI-negative subgroup (Figure \ref{fig:modelE_transitionDiagram_lowHBO}) is given by
\begin{equation}\label{eqn:modelE_transitionMatrix_HBOts}
M_{\text{D}_{\text{CADt}}} = 
 \left[
    \begin{array}{c|c|c|c}
      B_{00} & B_{01} & & \\
      \hline
      B_{10} & A_{1} & A_{2} & \\
      \hline
      & A_0 & A_{1} & \ddots \\
      \hline
      & & A_{0} & \ddots \\
      \hline
      & & & \ddots \\
    \end{array}
    \right]
\end{equation}
The $14\times14$ $A$ sub-matrices are defined below, where $\mathbb{0}_{12}$ denotes a $12\times12$ zero matrix, and $\mathbb{I}_{14}$ is a $14\times14$ identity matrix.

\begin{align*}\label{eqn:modelE_transitionMatrix_HBOts_As} 
 A_0 &= \scalemath{1}{\begin{pmatrix}
     (1-NPV)\mu_D    & NPV\mu_D    & \\
     (1-NPV)\mu_{ND} & NPV\mu_{ND} & \\
                     &             & \mathbb{0}_{12}
   \end{pmatrix}}, \\ \\
 A_2 &= \lambda_- \mathbb{I}_{14},
\end{align*}
\begin{strip}
\begin{align*}
 A_1&= \scalemath{0.85}{ \begin{pmatrix}[cc:c:c:c:c:c:c]
        * & & \mathbf{p}\lambda_{\text{em}} & & \mathbf{p}PPV\lambda_+ & & \mathbf{p}(1-PPV)\lambda_+ & \\
        & * & & \mathbf{p}\lambda_{\text{em}} & & \mathbf{p}PPV\lambda_+ & & \mathbf{p}(1-PPV)\lambda_+ \\[6pt]
        \hdashline[2pt/2pt]
        \mathbf{t}^{(1)} & & \mathbb{T}^{(1)}_3 & & & & & \\[6pt]
        \hdashline[2pt/2pt]
        & \mathbf{t}^{(1)} & & \mathbb{T}^{(1)}_4 & & & & \\[6pt]
        \hdashline[2pt/2pt]
        \mathbf{t}^{(2)} & & & & \mathbb{T}^{(2)}_5 & & &\\[6pt]
        \hdashline[2pt/2pt]
        & \mathbf{t}^{(2)} & & & & \mathbb{T}^{(2)}_6 & & \\[6pt]
        \hdashline[2pt/2pt]
        \mathbf{t}^{(3)} & & & & & & \mathbb{T}^{(3)}_7 &\\[6pt]
        \hdashline[2pt/2pt]
        & \mathbf{t}^{(3)} & & & & & & \mathbb{T}^{(3)}_8 
   \end{pmatrix}}, \numberthis
\end{align*}
\end{strip}

where, for a busy period $\textbf{B}_i$,
\begin{align*}
 \mathbf{p} = \begin{pmatrix}
    1 & 0
 \end{pmatrix}, \enspace
 \mathbf{t}^{(i)} = \begin{pmatrix}
    t_1^{(i)} \\[6pt] t_2^{(i)}  
 \end{pmatrix},  \enspace
  \mathbb{T}^{(i)}_k = \begin{pmatrix}
    * & t_{12}^{(i)} \\[6pt]
    0 & *
 \end{pmatrix}. \numberthis
\end{align*}
\\ \\ \\ \\ \\ \\
The boundary $B$ matrices, on the other hand, are

\begin{align*}\label{eqn:modelE_transitionMatrix_HBOts_Bs}
B_{00}&= \scalemath{1}{\begin{pmatrix}[c:c:c:c]
    -\sigma_1 & \mathbf{p}\lambda_{\text{em}} & \mathbf{p}PPV\lambda_+ & \mathbf{p}(1-PPV)\lambda_+ \\[6pt]
    \hdashline[2pt/2pt]
    \mathbf{t}^{(1)} & \mathbb{T}^{(1)}_2 & & \\[6pt]
    \hdashline[2pt/2pt]
    \mathbf{t}^{(2)} & & \mathbb{T}^{(2)}_3 & \\[6pt]
    \hdashline[2pt/2pt]
    \mathbf{t}^{(3)} & & & \mathbb{T}^{(3)}_4\\
   \end{pmatrix}}, \\ \\
 B_{01} &= \scalemath{1}{\begin{pmatrix}[cc:c:c:c]
    (1-NPV)\lambda_- & NPV\lambda_- & & & \\[6pt]
    \hdashline[2pt/2pt]
    & & \mathbb{Q} & & \\[6pt]
    \hdashline[2pt/2pt]
    & & & \mathbb{Q} & \\[6pt]
    \hdashline[2pt/2pt]
    & & & & \mathbb{Q}
   \end{pmatrix}}, \\ \\
 B_{10} &= \scalemath{1}{\begin{pmatrix}
     \mu_D    & \\
     \mu_{ND} & \\
              & \mathbb{0}_{12\times6}
   \end{pmatrix}}, \numberthis
\end{align*}

where 

\begin{align*}
  \mathbb{Q} = \scalemath{0.9}{\begin{pmatrix}
    (1-NPV)\lambda_-  & 0 & NPV\lambda_- & 0 \\[6pt]
    0 & (1-NPV)\lambda_-  & 0 & NPV\lambda_-
 \end{pmatrix}}. \\ \\ \numberthis
\end{align*}

\end{document}